\documentclass[fleqn,usenatbib]{mnras}

\usepackage{newtxtext,newtxmath}

\usepackage[T1]{fontenc}
\usepackage{ae,aecompl}

\usepackage{subfigure}
\usepackage{float}
\usepackage{amsmath}
\usepackage{etoolbox}
\pretocmd{\abstractname}{\newpage}{}{}

\usepackage{graphicx}	
\usepackage{amsmath}	
\usepackage{amssymb}	

\title[SCUBA-2 Reddened Quasar Survey]{A SCUBA-2 850$\mu$m Survey of Heavily Reddened Quasars at $z\sim2$}

\author[C. F. Wethers et al.]{
Clare F. Wethers$^{1, 2},$\thanks{E-mail: clweth@utu.fi}
Manda Banerji$^{2,3}$,
Paul C. Hewett$^{2}$,
Gareth C. Jones$^{4,3}$ 
\\
$^{1}$ Finnish Centre for Astronomy with ESO (FINCA), Vesilinnantie 5, FI-20014 University of Turku, Finland \\
$^{2}$Institute of Astronomy, University of Cambridge, Madingley Road, Cambridge, CB30HA, UK\\
$^{3}$Kavli Institute for Cosmology, University of Cambridge, Madingley Road, Cambridge, CB3 0HA, UK \\
$^{4}$Battcock Centre for Experimental Astrophysics, Cavendish Laboratory, Cambridge CB3 0HE, UK \\
}

\date{Accepted XXX. Received YYY; in original form ZZZ}

\pubyear{2015}

\begin{document}
\label{firstpage}
\pagerange{\pageref{firstpage}--\pageref{lastpage}}
\maketitle

\begin{abstract}
We present new 850$\mu$m SCUBA-2 observations for a sample of 19 heavily reddened Type-I quasars at redshifts $z\sim$2 with dust extinctions of A$_{\rm{V}} \simeq 2-6$ mag. Three of the 19 quasars are detected at $>$3$\sigma$ significance corresponding to an 850$\mu$m flux-limit of $\gtrsim$4.8 mJy. Assuming the 850$\mu$m flux is dominated by dust heating due to star formation, very high star formation rates (SFR) of $\sim$2500-4500 M$_\odot$ yr$^{-1}$ in the quasar host galaxies are inferred. Even when considering a large contribution to the 850$\mu$m flux from dust heated by the quasar itself, significant SFRs of $\sim$600-1500 M$_\odot$ yr$^{-1}$ are nevertheless inferred for two of the three detected quasars. We stack the remaining 16 heavily reddened quasars and derive an average 3$\sigma$ upper limit on the SFRs in these quasar host galaxies of $<$880 M$_\odot$ yr$^{-1}$. The number counts of sub-mm galaxies in the total survey area (134.3\,arcmin$^2$) are consistent with predictions from blank-field surveys. There are, however, individual quasars where we find evidence for an excess of associated sub-mm galaxies. For two quasars, higher spatial resolution and spectroscopic ALMA observations confirm the presence of an excess of sub-mm sources. We compare the 850$\mu$m detection rate of our quasars to both unobscured, ultraviolet luminous quasars as well as the much more obscured population of mid-infrared luminous Hot Dust Obscured Galaxies (HotDOGs). When matched by luminosity and redshift, we find no significant differences in the 850$\mu$m flux densities of these various quasar populations given the current small sample sizes. 
\end{abstract}




\section{Introduction}

Understanding the interactions between quasars and their host galaxies is critical for constraining models of galaxy evolution. In the local universe, tight correlations have been observed between the mass of the central super-massive black hole (SMBH) and that of the stellar bulge in galaxies \citep{magorrian98,kormendy13,gultekin09}, leading to the idea that black holes and galaxies co-evolve. Both star formation and black-hole accretion also peak at z $\sim$ 2 \citep[e.g.][]{madau14,aird15}, further suggesting a link between these two quantities. Whilst the co-evolution of galaxies and quasars is now widely accepted, the exact processes that shape this joint evolution remains an active area of research. Tidal interactions caused by major mergers can drive gas into the central regions of the galaxy, triggering star formation \citep{sanders88,barnes92,veilleux02,hopkins06}. As a result of the increased gas supply, the active galactic nucleus (AGN) is able to accrete close to the Eddington limit, appearing highly luminous yet dust-obscured as material from the remnant starburst is blown out from the galaxy. This so-called \emph{blowout} phase \citep{dimatteo05} is postulated to mark the transition between starburst galaxies and ultraviolet-luminous quasars and thus may represent a key phase in the lifetime of a quasar. Studying the connection between dust obscuration, black-hole accretion and star formation in quasars undergoing this transition therefore remains an important test of galaxy evolution models.

In \cite{wethers18}, we explored the link between star formation as traced by rest-frame ultraviolet (UV) emission and quasar luminosity in a sample of luminous, yet dust-obscured quasars from \citep{banerji12,banerji15} thought to be seen during the \emph{blowout} phase. These so-called heavily reddened quasars have dust extinctions of A$_{\rm{V}} \simeq 2-6$ mags, comparable to what is seen in the sub-millimetre galaxy (SMG) population \citep{takata06}. Independent observations of the narrow-line region in these quasars \citep{temple19} as well as ALMA follow-up \citep{banerji17, banerji18} suggests that the obscuring dust is likely distributed on galaxy wide scales. In our previous study, we found that these reddened quasars reside in prodigiously star-forming hosts (SFR$_{\rm{UV}}$ $\sim$ 100 M$_{\odot}$yr$^{-1}$), with the most luminous quasars residing in the most actively star-forming host galaxies. However, the rest-frame UV wavelengths considered in \cite{wethers18} trace only the unobscured component of the star formation in the host -- potentially accounting for only a small fraction of the total. Long-wavelength observations in the far infrared and at sub-millimetre (sub-mm) wavelengths, tracing cool dust emission, offer the opportunity to study the obscured components of star formation in these systems (e.g. \citealt{banerji14, banerji17}).  Early studies of UV-luminous high-redshift quasars at sub-mm wavelengths with SCUBA found no strong correlation between the sub-mm and optical luminosity of the quasar at $z\sim2$ \citep{priddey03} or $z>4$ \citep{isaak02}. More recently, studies of the far infrared-derived star-formation rates in larger samples of AGN covering a wider range in AGN luminosities and using a combination of data from \textit{Herschel} and SCUBA-2 have been undertaken. The results variously reveal either no trend in star-formation rate with quasar luminosity (e.g. \citealt{harrison12}) or, potentially, a correlation between quasar luminosity and star formation rate at the highest luminosities, particularly for populations of obscured quasars (e.g. \citealt{rosario12, chen13, banerji15a, stanley17}). Thus, the link between star formation, quasar luminosity and dust obscuration in high-luminosity, high-redshift quasars still remains unclear.

In this paper we seek to directly test whether the dust obscuration towards the quasar continuum seen in our previously studied population of $z\sim2$ heavily reddened quasars \citep{banerji12, banerji15, wethers18} is directly linked to the level of obscured star formation in the quasar host galaxy. To this end we have conducted a flux-limited 850$\mu$m survey with SCUBA-2 of our heavily reddened quasar population. The paper is structured as follows. Section~\ref{sec:data} outlines the sample of heavily reddened quasars and provides detail of the SCUBA-2 850$\mu$m observations. Section~\ref{sec:results} presents the main results of our study including inferences on obscured star formation in the quasar host galaxies from the 850$\mu$m photometry and a study of the environments of our heavily reddened quasars using the SCUBA-2 maps. Section~\ref{sec:discussion} discusses our results in the context of independent studies of similar objects, with our key findings summarised in Section~\ref{sec:conclusions}. Throughout this work, we adopt a flat $\Lambda$CDM cosmology with $H_{0}$ = 70 km s$^{-1}$ Mpc$^{-1}$, $\Omega_{M}$ = 0.3 and $\Omega_{\Lambda}$ = 0.7.

\section{Data}
\label{sec:data}

The parent population of targets studied in this work is a sample of heavily reddened (A$_{\rm{V}}$ $\simeq $ 2 - 6 mags), high-luminosity (L$_{\rm{bol}} \sim$ 10$^{47}$ erg/s) quasars at redshifts of 0.7 $<$ z $<$ 2.7. The photometric selection and spectroscopic confirmation of these quasars has been presented in \cite{banerji12,banerji13,banerji15} and \cite{temple19} but is summarised briefly below. 

\subsection{Heavily Reddened Quasars}
Heavily reddened quasar candidates were identified in wide-field near-infrared imaging surveys such as the UKIDSS Large Area Survey (ULAS; \citealt{lawrence07}),  VISTA Hemisphere Survey (VHS; \citealt{mcmahon13}) and VIKING \citep{edge13}. Targets were required to appear as point-sources in the $K$-band, with \emph{K}$_{\rm{Vega}} <$ 18.4\footnote{Magnitudes were calculated within a 1\,arcsecond radius aperture (\textit{apermag3}) and include an aperture correction appropriate for point sources.} to exclude galaxies and isolate the sample of high redshift, near-infrared-luminous quasars. A colour cut of (\emph{J}-\emph{K})$_{\rm{Vega}}>$~2.5 was applied to select targets with extremely red near-infrared colours, indicating significant dust reddening (A$_{\rm{V}}$ $\simeq$ 2 - 6 mags). Spectroscopic confirmation made use of the VLT-SINFONI and Gemini-GNIRS spectrographs and has now resulted in a sample of 63 heavily reddened quasars at 0.7 $<z<$ 2.7. From these 63 quasars we identified a high-redshift sample with $z>1.4$, which were also observable from Mauna Kea (Dec $>-17\deg$). The 19 quasars thus identified, most of which are at $z>2$,  were targeted with SCUBA-2. With extinction-corrected bolometric luminosities of L$_{\rm{bol}}\sim$10$^{47}$ergs$^{-1}$ and M$_{\rm{BH}}$ $\sim$10$^{9-10}$ M$_\odot$ \citep{banerji15}, the objects are among the most luminous and massive accreting SMBHs known at z$\sim$2.

\subsection{SCUBA2 Observations and Data Reduction}
\label{sec:reduction}
SCUBA-2 is a dual wavelength camera, operating at 450$\mu$m for the short waveband and 850$\mu$m for the long waveband, with diffraction-limited beam full-width half maxima (FWHM) of approximately 9.8- and 14.6\,arcsec respectively. Each of the two focal planes comprise four sub-arrays of 1280 bolometers, totalling 5120 pixels in each plane \citep{holland13}.

The 19 quasar targets were observed with SCUBA-2 on the 15m \textit{James Clerk Maxwell Telescope} (JCMT) on Mauna Kea, Hawaii, between 2017 March and May. Each target was observed in the `CV Daisy' observing mode, which is suitable for point-like sources or sources with structure on scales $<$ 3\,arcmin. The CV Daisy observing-mode produces a 12\,arcmin diameter map, with the deepest coverage in the central 3\,arcmin diameter region, where all four sub-arrays overlap \citep{holland13}. Details of the 19 reddened quasars observed with SCUBA-2 are given in Table~\ref{tab:observations}. The exposure time for each observation ranges between 25 and 37 minutes, reaching a 1$\sigma$ noise level of $\sim1.1-2.6$ mJy.

The 850$\mu$m SCUBA-2 observations are reduced with the \textsc{starlink} SubMillimetre User Reduction Facility (\textsc{smurf}) data reduction package. We used the `\textsc{reduce\_scan\_faint \_point\_sources}' recipe, which uses a `blank field' configuration suitable for point sources with low signal-to-noise ratio \citep{chapin13}. Initial pre-processing is carried out by the \textsc{smurf} software, which cleans the data by modelling the signal contributions from each bolometer, removing unwanted atmospheric emission and re-gridding the data to produce a science-quality image. Each of these science-quality images are then combined as a mosaic, centred on the sky position of the source, by the PIpeline for Combining and Analysing Reduced Data (\textsc{picard}). Finally, a beam-matched filter is applied to the resulting mosaicked image, smoothing the image with a 15\,arcsec FWHM Gaussian and calibrating the resulting map with a flux conversion factor of 2.34 Jy pW$^{-1}$arcsec$^{-2}$ (appropriate for point sources). In this way, 850$\mu$m maps are created for each of the 19 reddened quasars in our sample. Finally, the astrometry and photometry of the resulting SCUBA-2 maps are checked via the reduction of the pointing and flux calibrators through the same pipeline. The noise level for each map is derived from the median of the error map of each observation (cropped to cover the same region as the 850$\mu$m signal maps) using the Kernel APplication PAckage (\textsc{kappa}) within the STARLINK software.

\begin{table*}
    \centering
    \caption{SCUBA-2 observation log for the 19 heavily-obscured quasars, including the 850$\mu$m flux densities and associated uncertainties derived in Section~\ref{sec:results_phot}.}
    \label{tab:observations}
    \begin{tabular}{llrcccccc} 
        \hline
        Name & RA & DEC & \multicolumn{1}{p{2.0cm}}{\centering Observation \\ Date} & \multicolumn{1}{p{2.cm}}{\centering Exposure Time \\ (HH:MM:SS)} & \multicolumn{1}{p{2.cm}}{\centering S$_{850}$ $\pm$ $\sigma_{850}$ \\ (mJy)} & \multicolumn{1}{p{2.cm}}{\centering log$_{10}$[L$_{\rm{bol}}$] \\ (ergs$^{-1}$)} & Redshift \\
        \hline
        ULASJ0016-0038 & 4.0025   & -0.6498 & 2017.05.06 & 00:32:28 &  0.15 $\pm$1.73 & 46.9 & 2.194 \\
        ULASJ0041-0021 & 10.3041  & -0.3520 & 2017.05.06 & 00:32:32 &  0.74 $\pm$1.70 & 47.3 & 2.517 \\
        VHSJ1117-1528 & 169.2656  &-15.4749 & 2017.04.17 & 00:36:47 &  0.52 $\pm$2.64 & 47.6 & 2.427 \\
        VHSJ1122-1919 & 170.6018  &-19.3215 & 2017.04.17 & 00:36:47 & -0.04 $\pm$1.62 & 47.6 & 2.465 \\
        ULASJ1216-0313 & 184.1324 & -3.2264 & 2017.04.16 & 00:32:14 &  9.90 $\pm$1.82 & 48.3 & 2.576 \\
        ULASJ1234+0907 & 188.6147 &  9.1317 & 2017.04.16 & 00:30:40 &  2.46 $\pm$2.03 & 48.5 & 2.503 \\
        VHSJ1301-1624 & 195.3805  &-16.4150 & 2017.04.17 & 00:36:53 &  2.49 $\pm$1.62 & 47.4 & 2.140 \\
        VHSJ1350-0503 & 207.6552  & -5.0665 & 2017.04.18 & 00:33:45 &  0.71 $\pm$1.85 & 47.2 & 2.176 \\
        VHSJ1409-0830 & 212.3721  & -8.5163 & 2017.04.18 & 00:34:49 &  3.26 $\pm$1.70 & 47.1 & 2.300 \\
        ULASJ1455+1230 & 223.8375 & 12.5024 & 2017.05.04 & 00:31:10 &  2.53 $\pm$1.85 & 47.2 & 1.460 \\
        ULASJ1539+0057 & 234.7923 &  5.9638 & 2017.04.04 & 00:30:57 &  4.06 $\pm$1.53 & 48.2 & 2.658 \\
        VHSJ1556-0835 & 239.1571  & -8.5952 & 2017.04.17 & 00:33:55 &  1.69 $\pm$1.58 & 47.0 & 2.188 \\
        VHSJ2109-0026 & 317.3630  & -0.4497 & 2017.04.03 & 00:32:02 &  1.51 $\pm$1.59 & 46.9 & 2.344 \\
        VHSJ2143-0643 & 325.8926  & -6.7206 & 2017.03.30 & 00:33:57 &  0.68 $\pm$1.54 & 47.1 & 2.383 \\
        VHSJ2144-0523 & 326.2394  & -5.3881 & 2017.04.04 & 00:25:09 &  0.75 $\pm$1.10 & 46.7 & 2.152 \\
        ULASJ2200+0056 & 330.1036 &  0.9346 & 2017.04.17 & 00:32:27 &  5.92 $\pm$1.64 & 47.4 & 2.541 \\
        ULASJ2224-0015 & 336.0392 & -0.2566 & 2017.04.18 & 00:32:04 & -1.83 $\pm$2.00 & 47.0 & 2.223 \\
        ULASJ2315+0143 & 348.9843 &  1.7307 & 2017.05.05 & 00:32:32 &  6.43 $\pm$1.58 & 47.5 & 2.560\\
        VHSJ2355-0011 & 358.9394  & -0.1893 & 2017.05.06 & 00:32:25 &  3.66 $\pm$1.72 & 47.2 & 2.531 \\
        \hline
    \end{tabular}
\end{table*}

\section{Results}
\label{sec:results}

\subsection{850$\mu$m Photometry}
\label{sec:results_phot}

\begin{figure*}
	\centering  
    \subfigure[ULASJ1216-0313]{\includegraphics[trim= 100 50 100 40 ,clip,width=0.3\textwidth]{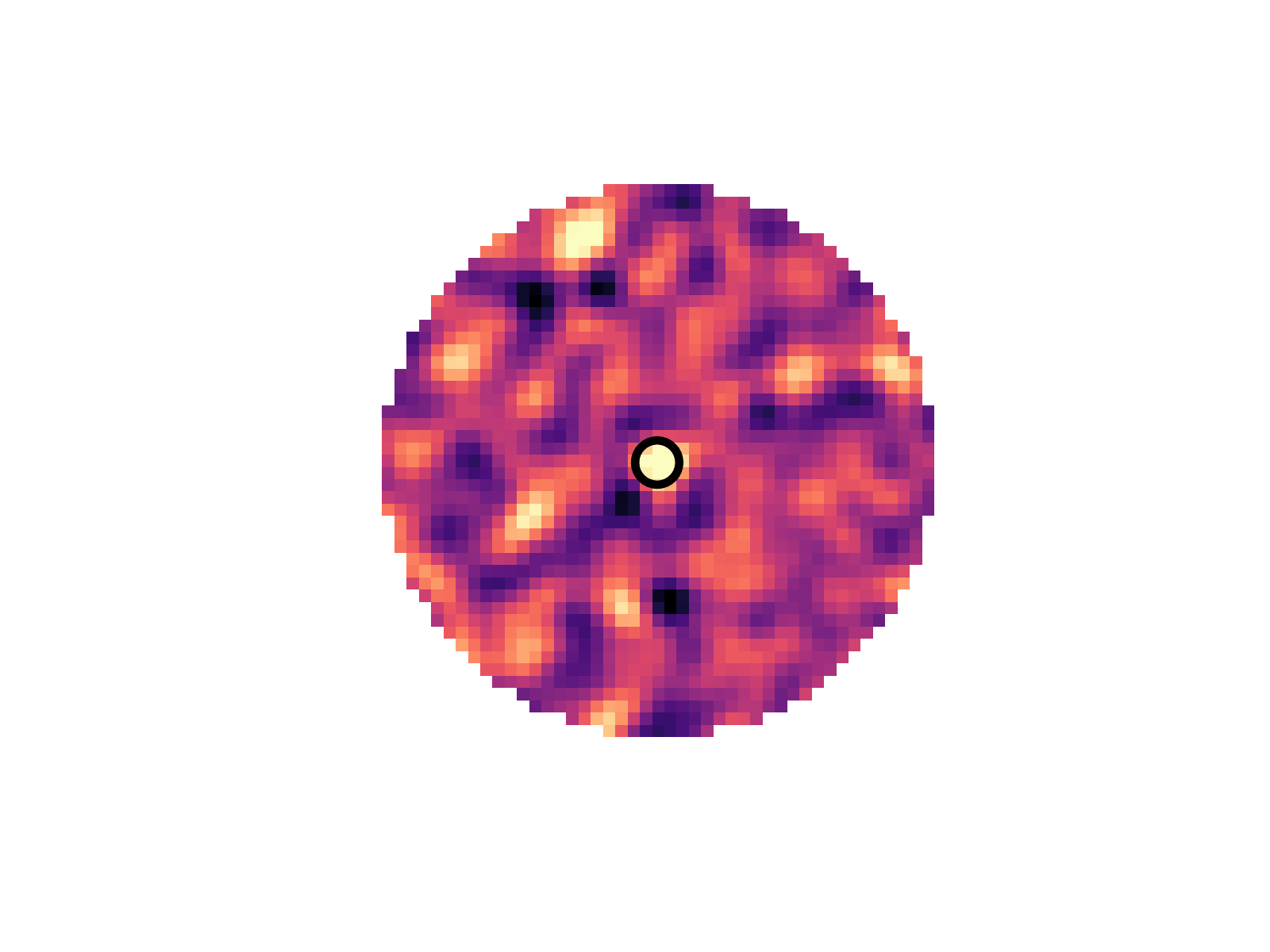}}
    \subfigure[ULASJ2200+0056]{\includegraphics[trim= 100 50 100 40 ,clip,width=0.3\textwidth]{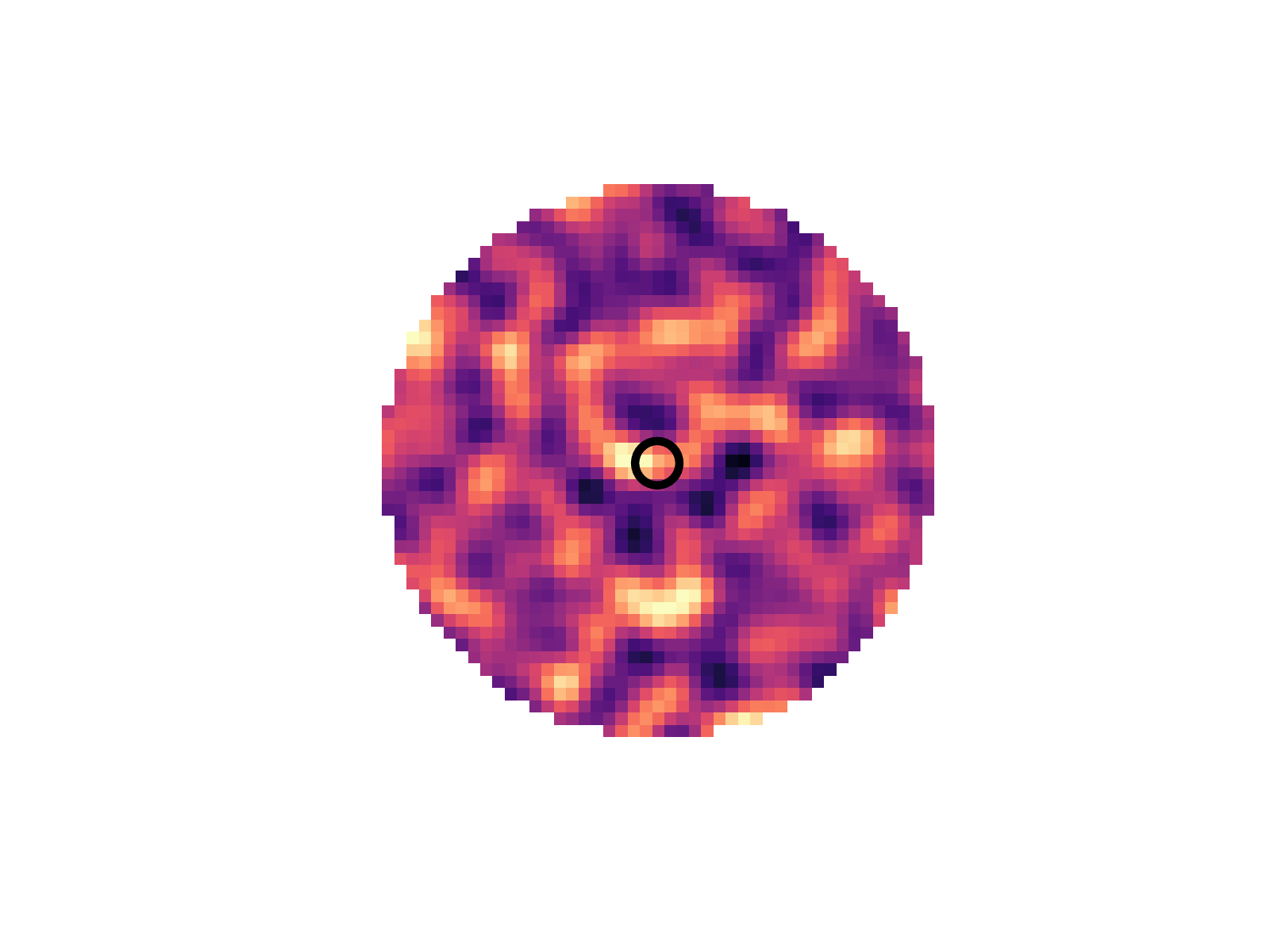}}
	\subfigure[ULASJ2315+0143]{\includegraphics[trim= 100 50 100 40 ,clip,width=0.3\textwidth]{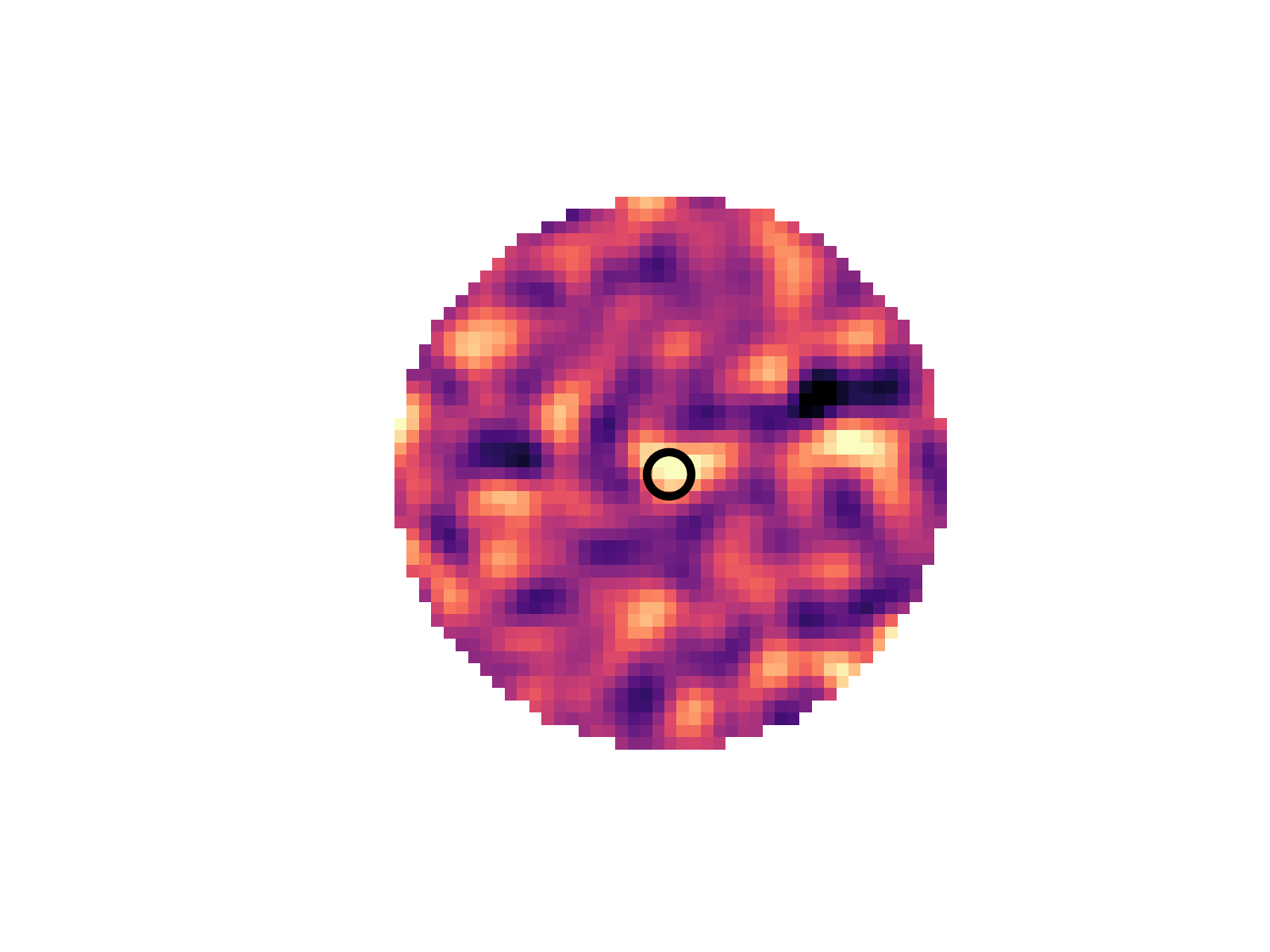}}
	\subfigure{\label{fig:cbar}\includegraphics[trim= 0 100 0 0 ,clip,width=0.9\textwidth]{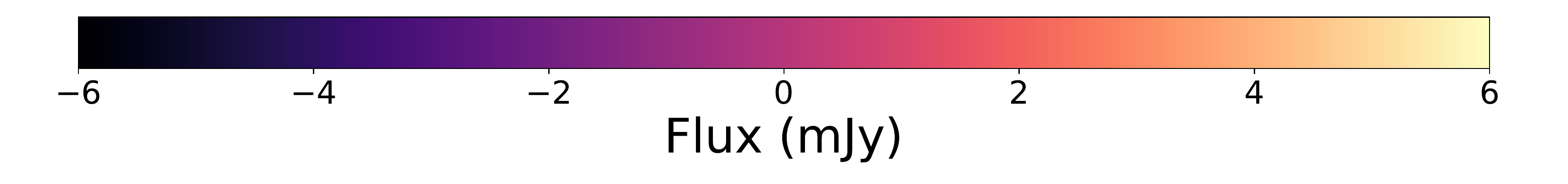}}
\caption{850$\mu$m SCUBA-2 maps for the three sources detected at a significance $>$3$\sigma$ - ULASJ1216-0313, ULASJ2200+0056 and ULASJ2315+0143. Black circle denotes a 14.6-arcsec diameter aperture centred on the SCUBA-2 position of the target source. North is up, East is to the left.}
\label{fig:detect}
\end{figure*}

From the reduced SCUBA-2 850$\mu$m maps, we calculate peak flux densities for the 19 quasar targets using a 14.6\,arcsec diameter aperture, centred on the quasar, which matches the beam size of SCUBA-2. These values are given in Table~\ref{tab:observations} along with the corresponding 1$\sigma$ noise on each measurement. Of the 19 quasars, only three sources are detected at $>$3$\sigma$ confidence. These are ULASJ1216-0313, ULASJ2200+0056 and ULASJ2315+0143 and the corresponding 850$\mu$m maps can be seen in Fig.~\ref{fig:detect}. We note that the peak sub-mm flux for ULASJ2200+0056 appears to be spatially offset from the quasar centroid. Nevertheless we class this source as a detection as the peak sub-mm flux lies within one SCUBA-2 beam from the quasar position. 

\begin{figure}
    \centering  
    \includegraphics[trim=50 20 30 30 ,clip,width=0.45\textwidth]{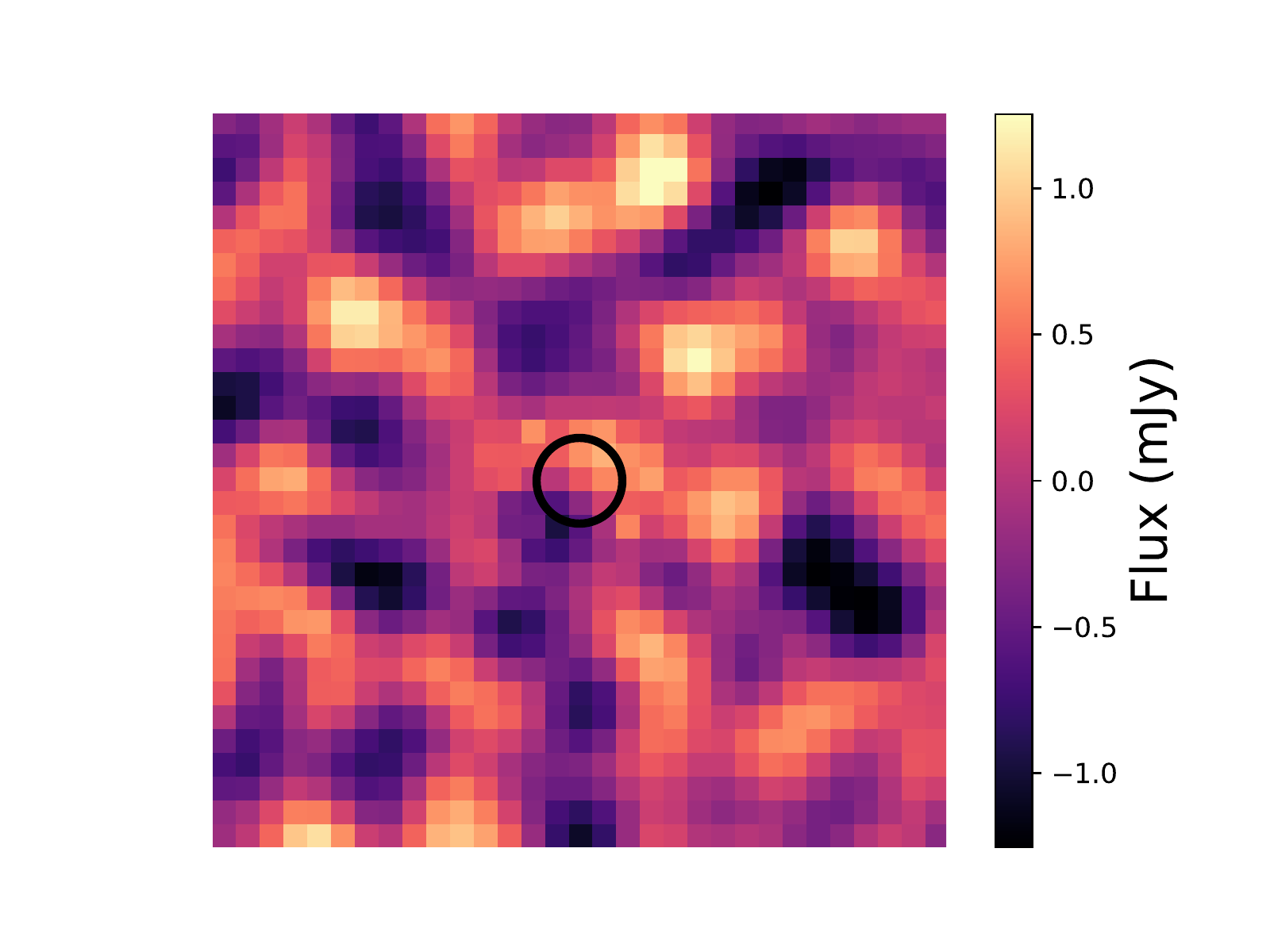}  
    \caption{Inverse variance weighted stack for the 16 undetected targets in our sample. Peak flux of 0.85 $\pm$ 0.41 mJy ($\sim$2.0$\sigma$) within a 14.6 arcsec SCUBA-2 beamsize.}
    \label{fig:stack}
\end{figure}

Fourteen of the remaining 16 quasar targets, which lie below the 3$\sigma$ detection limit nevertheless have a positive 850$\mu$m peak flux density within a beam-sized aperture centred on the near-infrared position of the quasar, with a significance of $\sim$1.1$\sigma$. To explore the average sub-mm properties of our sample we stack all 16 undetected sources, which all lie at z$\sim$2. An inverse variance weighting is applied to the stack to account for noise variations between objects i.e.

\begin{equation}
S_{\rm{ij}} = \frac{\sum_{\rm{k=1}}^{\rm{N}} P_{\rm{ij}}^{\rm{k}} \times f_{\rm{ij}}^{\rm{k}} /  (\sigma_{\rm{ij}}^{\rm{k}} \times \sigma_{\rm{ij}}^{\rm{k}})}{\sum_{\rm{k=1}}^{\rm{N}} P_{\rm{i,j}}^{\rm{k}} /  (\sigma_{\rm{ij}}^{\rm{k}} \times \sigma_{\rm{ij}}^{\rm{k}})},
\end{equation}

where $i,j$ denote the pixel coordinates with respect to the catalogued quasar position. \noindent $P_{ij}$ is the response function of SCUBA-2 at 850$\mu$m, taken to be the PSF for each observation produced by the \textsc{starlink} reduction software, $f_{ij}$ are the flux densities for each of the $N$ sources in our sample and $\sigma_{ij}$ are the corresponding noise maps, provided independently from the combined signal + noise maps during the reduction (Sec.~\ref{sec:reduction}). The resulting stacked image is shown in Fig.~\ref{fig:stack}, from which we derive a peak flux density of 0.85 $\pm$ 0.41 mJy. The uncertainty on the stacked flux has been derived by looking at the standard deviation of the pixels over a blank portion of the stacked image prior to the PSF weighting. The stacked image therefore does not show a significant detection. However the average 850$\mu$m flux density derived for this small sample is consistent with \cite{banerji15a}, who have studied the average 850$\mu$m properties of a much larger sample of 699 X-ray luminous AGN and find a significant stacked flux density of 0.71$\pm$0.08 mJy.

\subsection{Dust-Obscured Star Formation}
\label{sec:sfrs}

Based on the photometry in Table~\ref{tab:observations}, we now derive star formation rates for the three sources in our sample detected at 850$\mu$m - ULASJ1216-0313, ULASJ2200+0056 and ULASJ2315+0143 - as well as an upper limit on the star formation rate based on the stacked non-detections (Fig.~\ref{fig:stack}). We assume for now that the cold dust component traced by the 850$\mu$m flux density at these redshifts is being heated entirely by star formation although see \citet{symeonidis17} and Section \ref{sec:quasarheating} for an alternative explanation. The far-infrared to sub-mm spectral energy distributions (SEDs) of our quasars can then be modelled using a single grey-body template, characterised by three parameters; the dust temperature, $T$, the emissivity index, $\beta$ and the overall normalisation, $f_{\rm{GB}}$, such that

\begin{equation}
S_\nu (\rm{mJy}) = f_{\rm{GB}} \times (1 - e^{\tau(\nu)}) \times \frac{\nu^3}{e^{\frac{h\nu}{kT}}-1},
\label{eqn:Sv_full}
\end{equation}

\noindent
where the optical depth, $\tau(\nu)$, is given as a function of $\nu_0$ - the frequency at which $\tau(\nu)$ = 1, i.e.
\begin{equation}
\tau(\nu) = \left(\frac{-\nu}{\nu_{0}} \right) ^{\beta}.
\end{equation}

\begin{figure}
	\centering  
    \subfigure[ULASJ1216-0313]{\includegraphics[trim= 15 0 30 40 ,clip,width=0.45\textwidth]{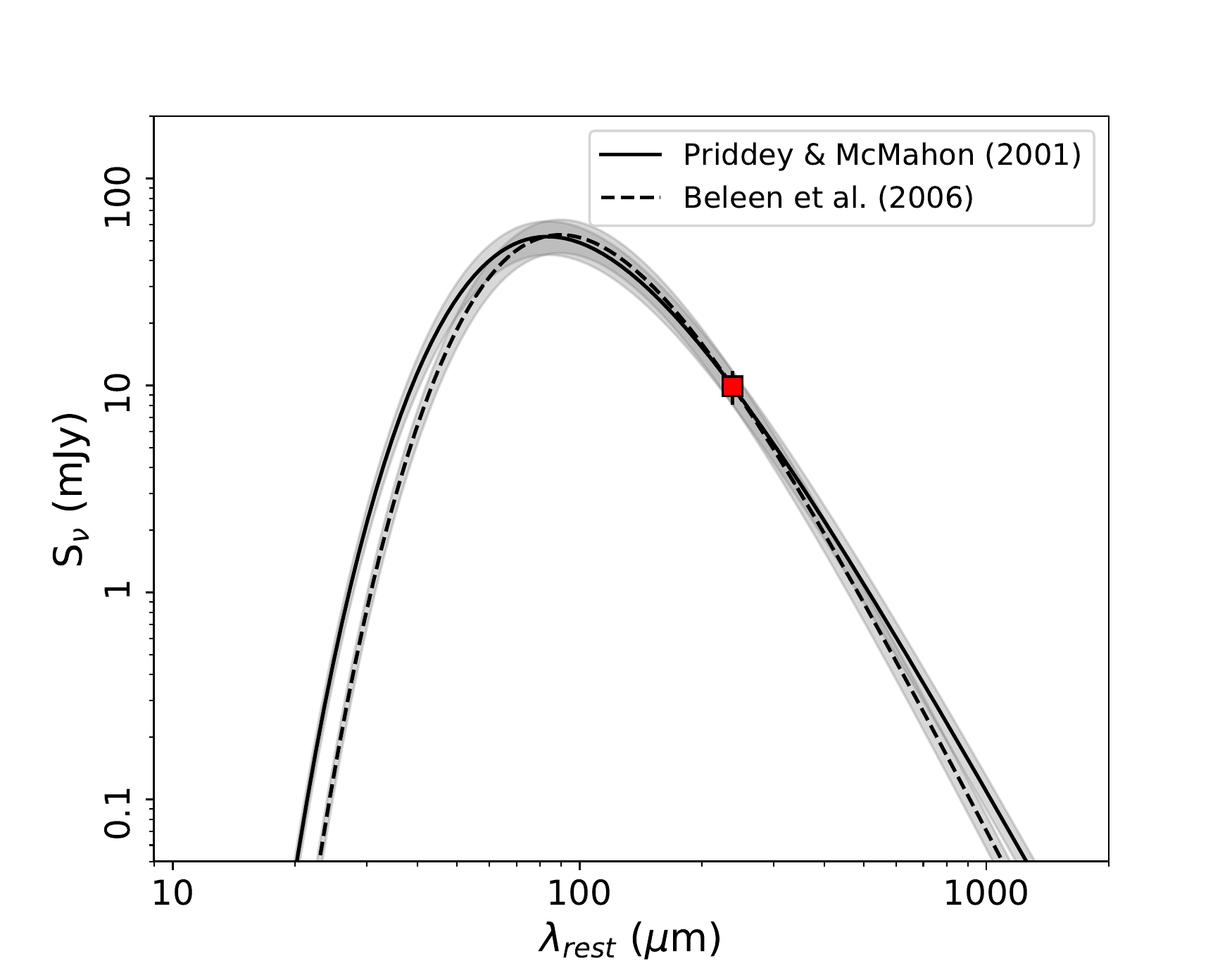}}
    \subfigure[ULASJ2200+0056]{\includegraphics[trim= 15 0 30 40 ,clip,width=0.45\textwidth]{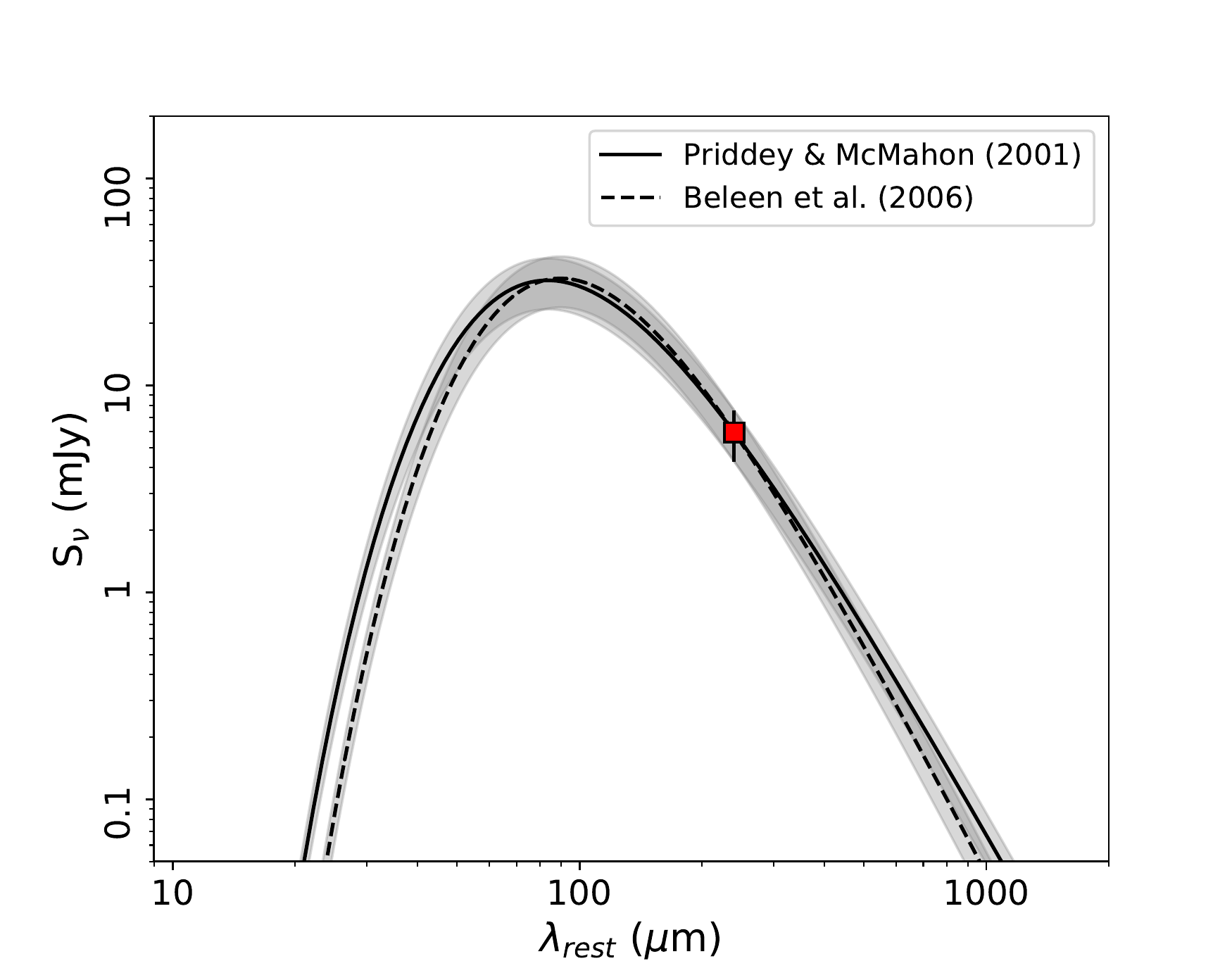}}
	\subfigure[ULASJ2315+0143]{\includegraphics[trim= 15 0 30 40 ,clip,width=0.45\textwidth]{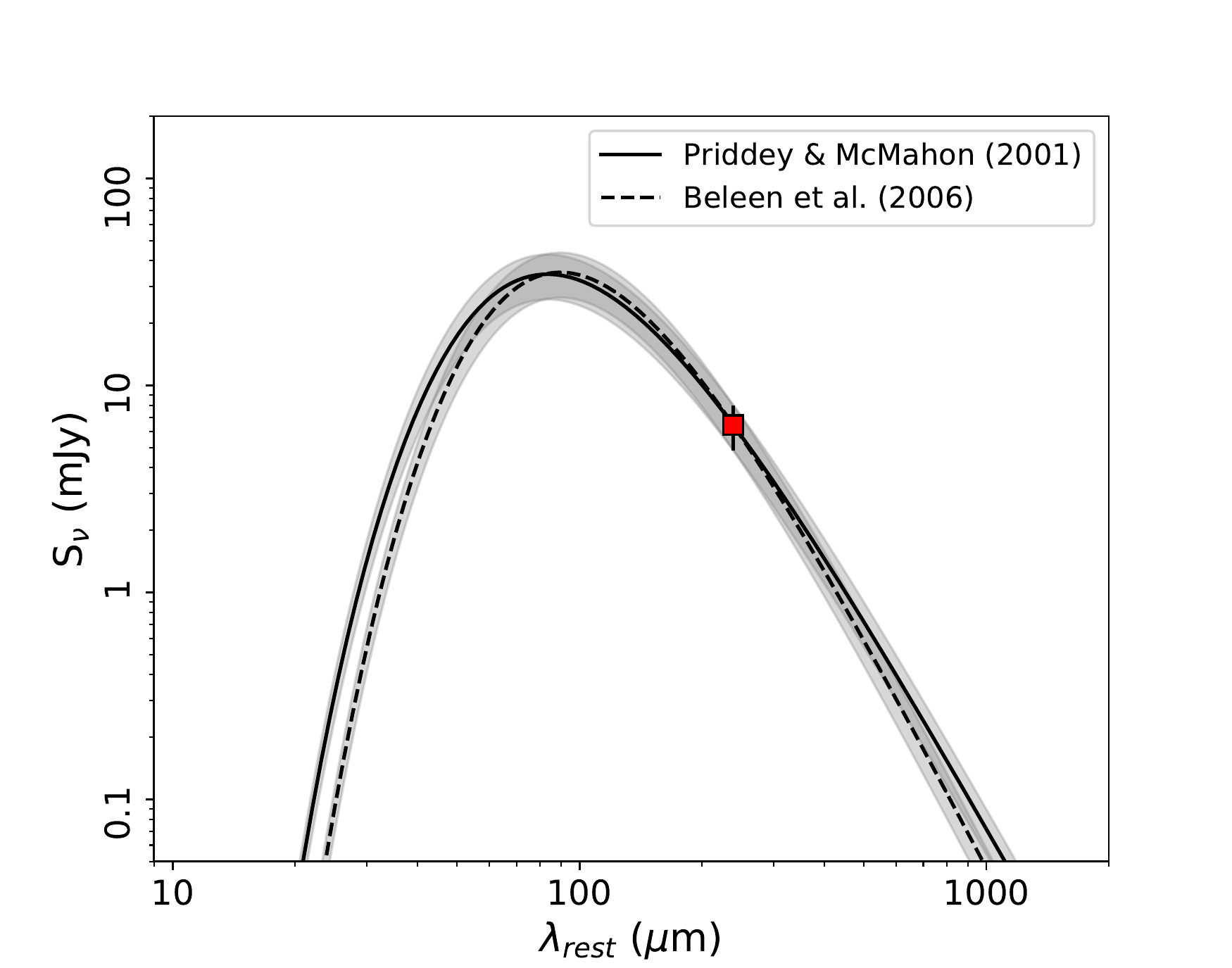}}
\caption{Grey-body fit to the 850$\mu$m photometry of ULASJ1216-0313 (a), ULASJ2200+0056 (b) and ULASJ2315+0143 (c) assuming (i) T=47K, $\beta$=1.6 \citep{priddey01} \emph{(solid)} and (ii) T=41K, $\beta$=1.95 \citep{beelen06} \emph{(dashed)}. Shaded regions represent the 1$\sigma$ uncertainty on each fit.}
\label{fig:greybody1}
\end{figure}

In high-redshift star-forming galaxies the dust temperatures lie in the range 20 $\lesssim$ $T$ $\lesssim$ 60 K (e.g. \citealt{casey12}), whereas in quasar host galaxies the dust is usually hotter, T$\sim$40-50K \citep{priddey01, beelen06}. Given the single photometric data point at 850$\mu$m available for our quasars we use two fixed combinations of $T$ and $\beta$ appropriate for high-redshift quasars: (i) T=47K, $\beta$=1.6 \citep{priddey01} and (ii) T=41K, $\beta$=1.95 \citep{beelen06}. Each greybody model is then fit in turn to the photometry of ULASJ1216-0313, ULASJ2200+0056 and ULASJ2315+0143, with the overall normalisation left as a free parameter in the fitting. The resulting grey-body fits are shown in Fig.~\ref{fig:greybody1}. The shaded regions in each of the greybody templates denote the 1$\sigma$ uncertainty in each fit, derived from fitting the upper and lower 1$\sigma$ bounds of the photometry.

From the resulting grey-body templates, we derive estimates for the far infrared luminosity:

\begin{equation}
L_{\rm{FIR}} = 4\pi D_{\rm{L}}^{2} \int_{\nu_{\rm{min}}}^{\nu_{\rm{max}}} S(\nu)  d\nu, 
\label{eqn:l_fir}
\end{equation}

\noindent where $D_{L}$ is the luminosity distance, $\nu_{min}$ and $\nu_{max}$ denote the frequency limits of the integral ($\nu_{min}$ = 1.0 THz; $\nu_{max}$ = 7.5 THz, corresponding to wavelengths 40$\mu$m $<$ $\lambda$ $<$ 300$\mu$m) and $S(\nu)$ is the best-fit grey-body. We use L$_{\rm{FIR}}$ - tracing the emission at 40$\mu$m $<$ $\lambda$ $<$ 300$\mu$m - rather than the total infrared luminosity in order to minimise contamination from dust heating by the quasar, which can significantly contribute to the flux at $\lambda$ $<$ 40$\mu$m. The key assumption is that the cooler dust, peaking at wavelengths $>$40$\mu$m, is predominantly heated by star formation. The SFR for each system is then derived using the relation from \cite{kennicutt12}:

\begin{equation}
SFR = 3.89\times10^{-44} \times L_{\rm{FIR}},
\label{l_sfr}
\end{equation}

\noindent where the SFR is given in units of M$_{\rm{\odot}}$yr$^{-1}$ and L$_{\rm{FIR}}$ is in units of erg/s. 

\begin{table}
    \centering
    \caption{SFRs for the targets detected at $>$3$\sigma$ confidence at 850$\mu$m, based on greybody template fitting with parameters from \citet{priddey01} (SFR$_{\rm{P}}$) and \citet{beelen06} (SFR$_{\rm{B}}$). For the 16 remaining non-detections, we derive 3$\sigma$ upper limits on the SFR based on each template.}
    \label{tab:sfrs}
    \begin{tabular}{lllcccccc} 
        \hline
        Name & SFR$_{\rm{P}}$ ( M$_{\rm{\odot}}$yr$^{-1}$) & SFR$_{\rm{B}}$ ( M$_{\rm{\odot}}$yr$^{-1}$) \\
        \hline
        ULASJ1216-0313 &  4450 $\pm$ 820 & 4080 $\pm$ 750 \\
        ULASJ2200+0056 &  2650 $\pm$ 730 & 2430 $\pm$ 670 \\
        ULASJ2315+0143 &  2880 $\pm$ 710 & 2650 $\pm$ 650 \\
        Stacked non-detections &  $<$ 880 & $<$ 830 \\
        \hline
    \end{tabular}
\end{table}

The SFRs based on the two different greybody assumptions for the three detected quasars and the 3$\sigma$ upper limits from the stacked non-detection are given in Table \ref{tab:sfrs}. In the case of the three 850$\mu$m detections, these SFRs are found to be very large - $\sim$2500-4500 M$_\odot$ yr$^{-1}$. Such high SFRs are not uncommon in the most luminous quasars detected at sub-mm wavelengths (e.g. \citealt{isaak02, priddey03, stacey18}) and even higher SFRs have been claimed in populations of hyperluminous infrared galaxies (HyLIRGs; \citealt{rowanrobinson18}). However, the three quasars detected at 850$\mu$m are also among the most luminous quasars in our sample and it is possible that the quasar itself could be contributing to the dust heating at 850$\mu$m, which would reduce the inferred SFRs. Later, in Section~\ref{sec:quasarheating}, we discuss the potential effects of this quasar contamination on our results. Two of the three quasars, ULASJ2200+0056 and ULASJ2315+0143 also showed evidence for residing in star-forming host galaxies based on our rest-frame ultraviolet detections of the quasar hosts in the Dark Energy Survey \citep{wethers18}. The ultraviolet-derived star formation rates for these two quasars are among the highest in our sample - SFR$_{\rm{UV}} \sim 200-400$M$_\odot$ yr$^{-1}$. Nevertheless, the sub-mm derived star formation rates are significantly higher, implying that a large fraction of the star formation in heavily reddened quasar hosts is obscured by dust. From the stacked non-detection of our remaining 850$\mu$m undetected quasars, we derive a 3$\sigma$ upper limit on the SFR of $\lesssim$850 M$_\odot$ yr$^{-1}$. We therefore note that while the majority of our heavily reddened quasars are undetected with SCUBA-2, this does not preclude significant levels of star formation in these systems. 

\subsubsection{Quasar Heating}

\label{sec:quasarheating}
\begin{figure}
	\centering
	\includegraphics[trim= 8 5 35 20 ,clip,width=0.45\textwidth]{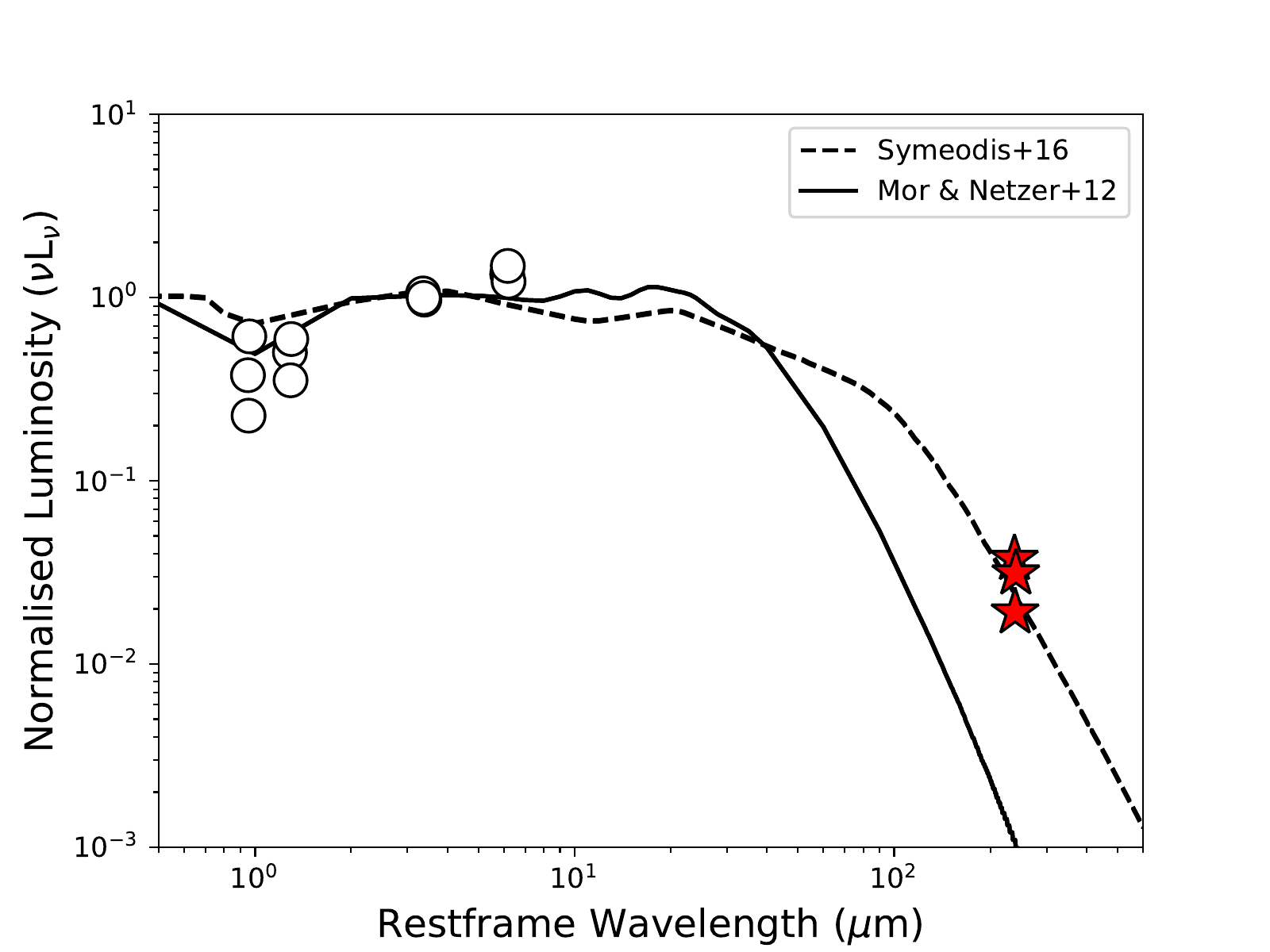}
\caption{Quasar SEDs from \citet{mor12} \textit{(solid line)} and \citet{symeonidis16} \textit{(dotted line)}, fitted to the WISE photometry of our detected quasar sample \textit{(white dots)} and normalised at 6$\mu$m. The SCUBA2 photometry \textit{(red stars)} is overlaid for reference.}
\label{fig:agn_sed}
\end{figure}

In Section~\ref{sec:sfrs}, we derive SFRs for the three quasars detected with SCUBA2, assuming the cold dust at 850$\mu$m is entirely heated by star formation. However, several studies suggest that a significant fraction of the emission at this wavelength may still be attributed to heating from the quasar, particularly in very luminous systems \citep[e.g.][]{mor12,symeonidis16,mullaney12,lyu17a,lyu17b}. To illustrate the level of uncertainty arising from the potential quasar contamination at 850$\mu$m, we make use of the quasar template provided in \cite{mor12}, based on the intrinsic SEDs of 115 nearby Type-1 AGN with luminosities L$_{5100}$ $\simeq$ 10$^{43.2}$-10$^{45.9}$ ergs$^{-1}$. The model is scaled to photometry from the Wide-field Infrared Survey Explorer (WISE) in four wavebands (W1;$\lambda$3.4$\mu$m, W2;$\lambda$4.6$\mu$m, W3;$\lambda$12$\mu$m and W4;$\lambda$22$\mu$m) and normalised at 6$\mu$m (Fig.~\ref{fig:agn_sed}). Based on this scaling, the contribution of the quasar to the 850$\mu$m flux of the three detected quasars (ULASJ1216-0313, ULASJ2200+0056 and ULASJ2315+0143) is estimated and subtracted from the original SCUBA2 photometry, and this new photometry is modelled with a modified blackbody curve following the methods outlined in Sec.~\ref{sec:sfrs}. Based on the model of \cite{mor12}, we find minimal quasar contamination at 850$\mu$m for all three of the quasars detected with SCUBA2, deriving quasar-subtracted SFRs within $\sim$5 per cent of the original estimates in Tab.~\ref{tab:sfrs} - SFR = 4320$\pm$710 (3970$\pm$750) M$_{\rm{\odot}}$yr$^{-1}$, 2560$\pm$730 (2360$\pm$670) M$_{\rm{\odot}}$yr$^{-1}$ and 2720$\pm$710 (2500$\pm$650) M$_{\rm{\odot}}$yr$^{-1}$ for ULASJ1216-0313, ULASJ2200+0056 and ULASJ2315+0143 respectively, based on the parameters in \cite{priddey03} \citep[and][]{beelen06}

Other studies however estimate much higher levels of quasar contamination at 850$\mu$m. \cite{symeonidis16}, for example, find that the contribution of quasar heating may be comparable to that of star formation out to $\lambda$ $<$ 1000$\mu$m. As we do not have enough sample specific information to constrain the exact level of quasar contamination in our sample, we additionally make use of the quasar SED in \cite{symeonidis16}, which includes the largest quasar contribution to the cool dust emission of quasar templates available in the literature and therefore provides the most conservative estimate of the SFR. Again, the quasar template is scaled to the WISE photometry of the three detected quasars (Fig.~\ref{fig:agn_sed}) and the quasar contribution at 850$\mu$m is subtracted from the SCUBA2 flux. Fitting a greybody curve to this new photometry returns significantly lower SFRs than derived in Sec.~\ref{sec:sfrs}. When accounting for the quasar contamination estimated by the \cite{symeonidis16} model, we derive SFR = 1570$\pm$820 (1440$\pm$750) and 690$\pm$730 (630$\pm$670) M$_{\rm{\odot}}$yr$^{-1}$ for ULASJ1216-0313 and ULASJ2200+0056 respectively, based on the parameters in \cite{priddey03} \citep[and][]{beelen06}, whilst all the 850$\mu$m flux observed in ULASJ2315+0143 can be accounted for by the \cite{symeonidis16} model.

Despite deriving SFRs a factor of $>$2.5 lower than those in Section~\ref{sec:sfrs}, in which all cool dust is attributed to star formation, both ULASJ1216-0313 and ULASJ2200+0056 continue to show evidence for prolific star formation. Based on the model in \cite{symeonidis16} however, we cannot rule out that the 850$\mu$m flux of ULASJ2315+0143 derives entirely from quasar heating. However, ALMA observations of ULASJ2315+0143 \citep{banerji18} have detected the dust continuum emission from this source at an observed frame wavelength of 1.2mm (rest-frame $\sim$330 $\mu$m) where contributions from quasar heating are expected to be negligible even when considering the most extreme quasar models. Full SED modelling of this source in \cite{banerji18} leads to an estimated star formation rate of 680 M$_{\rm{\odot}}$yr$^{-1}$. The source is additionally shown to be a close separation merger of two separate dusty galaxies. Thus at least some of the 850um emission in ULASJ2315+0143 is likely to be originating from dust heated by star formation. Whilst the effects of quasar contamination may significantly alter the SFRs derived for our quasar sample, we therefore maintain that the dusty quasars detected with SCUBA2 show evidence for prolific star formation.

\subsection{Reddened Quasar Environments}
\label{sec:environments}

Whilst the majority of our reddened quasar sample are undetected at 850$\mu$m, inspection of the maps reveals a number of serendipitous sub-mm detections within a 1.5 arcmin radius of the quasar targets, potentially indicating an environmental overdensity on scales $<$ 1Mpc. To investigate this and to identify any serendipitous detections, we look for $>$3$\sigma$ peaks in the signal-to-noise ratio maps. Any sources lying at the edge of the maps were omitted from the source counts. 

\begin{figure*}
	\centering  
	\subfigure[ULASJ0016-0038]{\includegraphics[trim= 120 70 115 65 ,clip,width=0.23\textwidth]{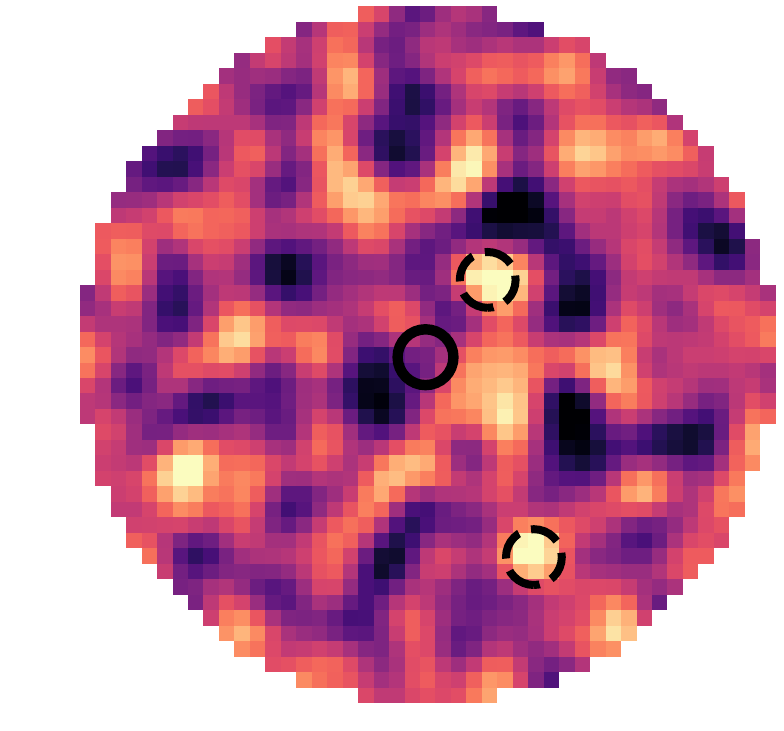}}
    \subfigure[ULASJ0041-0021]{\includegraphics[trim= 120 70 115 65 ,clip,width=0.23\textwidth]{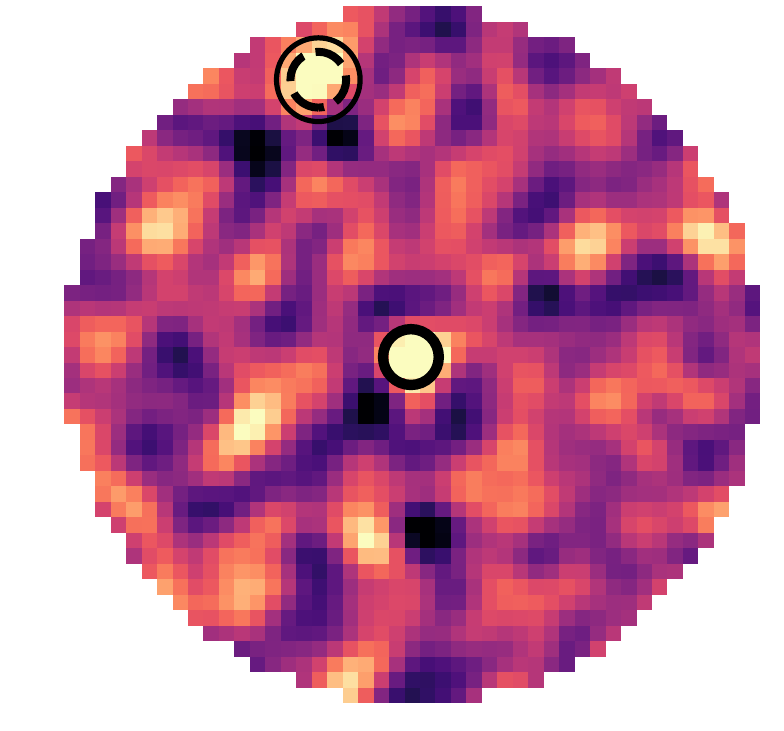}}
	\subfigure[VHSJ1117-1528]{\includegraphics[trim= 120 70 115 65 ,clip,width=0.23\textwidth]{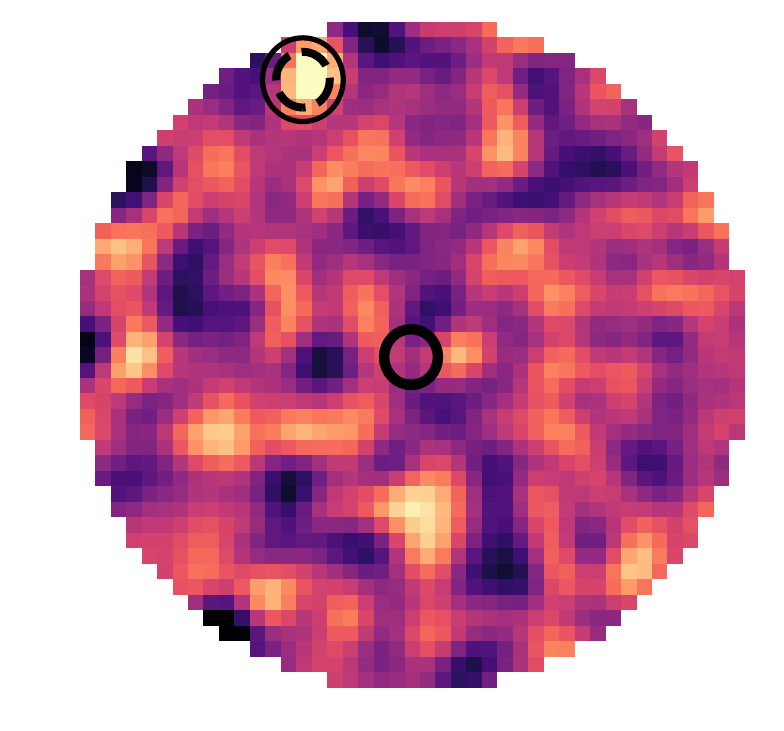}}
    \subfigure[VHSJ1122-1919]{\includegraphics[trim= 120 70 115 65 ,clip,width=0.23\textwidth]{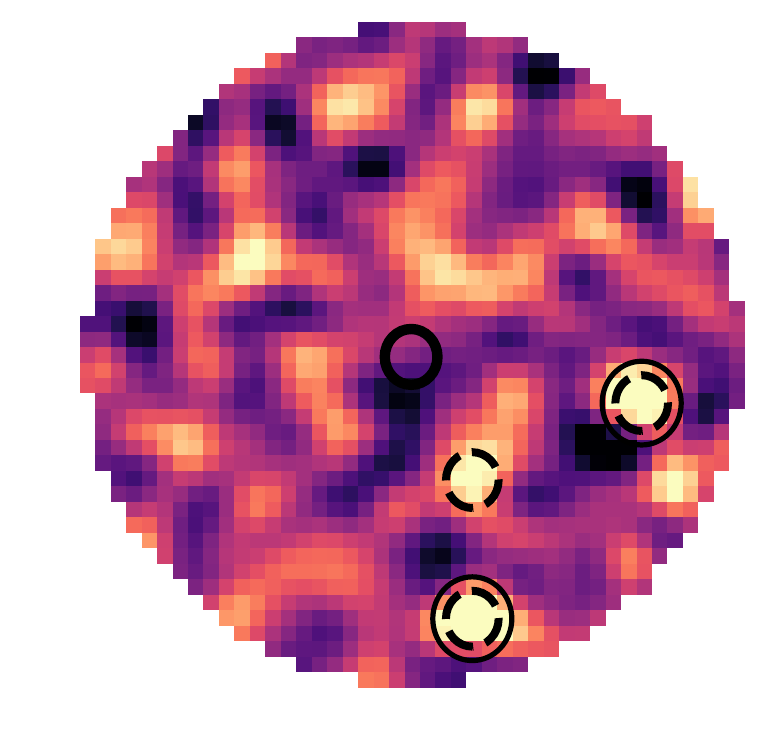}}
	\subfigure[ULASJ1216-0313]{\includegraphics[trim= 120 70 115 65 ,clip,width=0.23\textwidth]{Figs/ULASJ1216_serendipitous_pos.pdf}}
	\subfigure[ULASJ1234+0907]{\includegraphics[trim= 120 70 115 65 ,clip,width=0.23\textwidth]{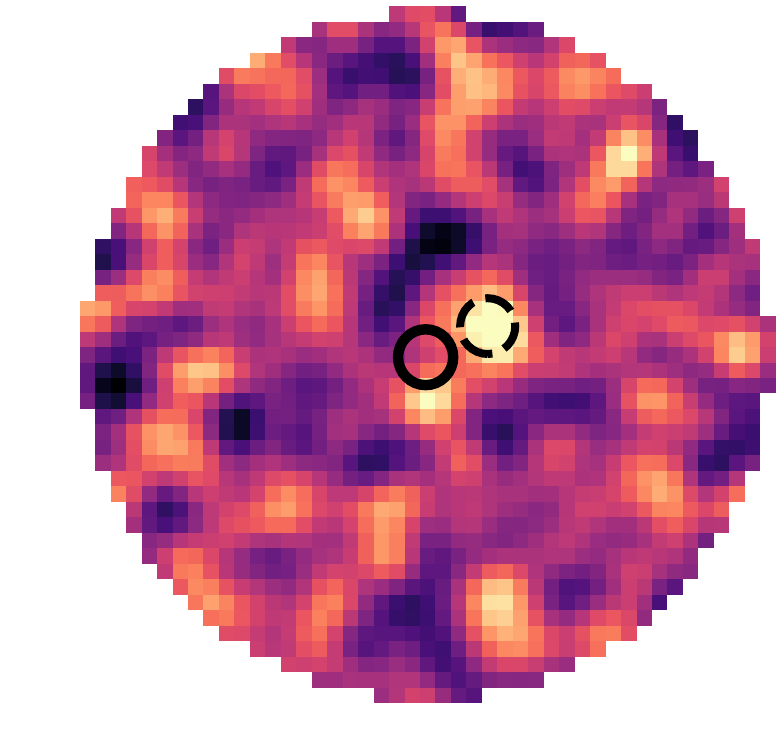}}
	\subfigure[VHSJ1301-1624]{\includegraphics[trim= 120 70 115 65 ,clip,width=0.23\textwidth]{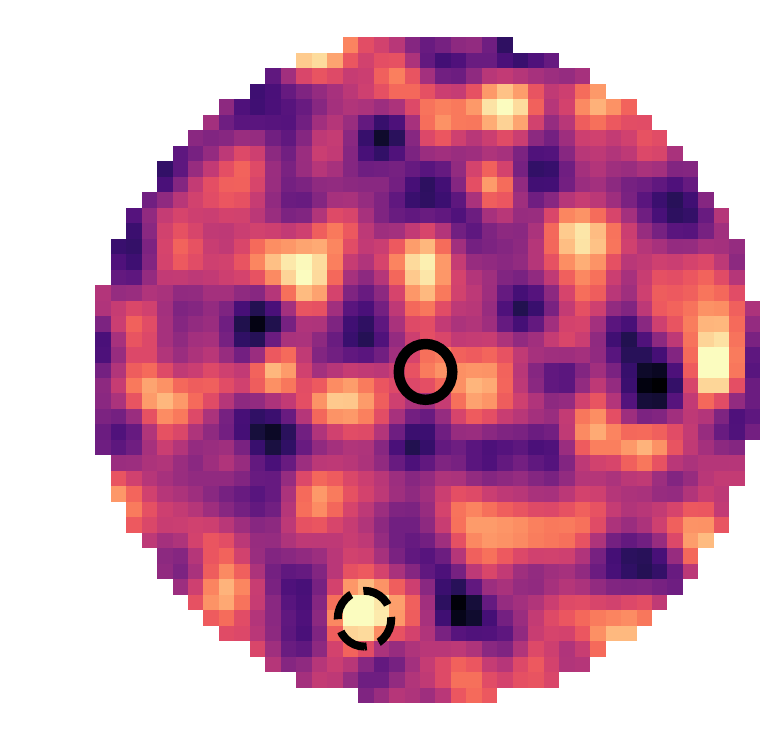}}
    \subfigure[VHSJ1350-0503]{\includegraphics[trim= 120 70 115 65 ,clip,width=0.23\textwidth]{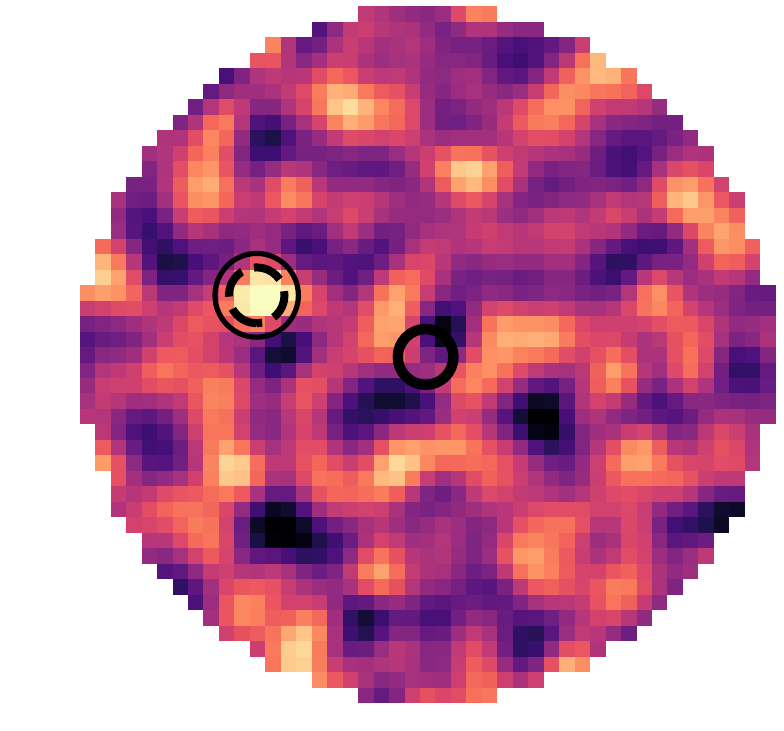}}
	\subfigure[VHSJ1409-0830]{\includegraphics[trim= 120 70 115 65 ,clip,width=0.23\textwidth]{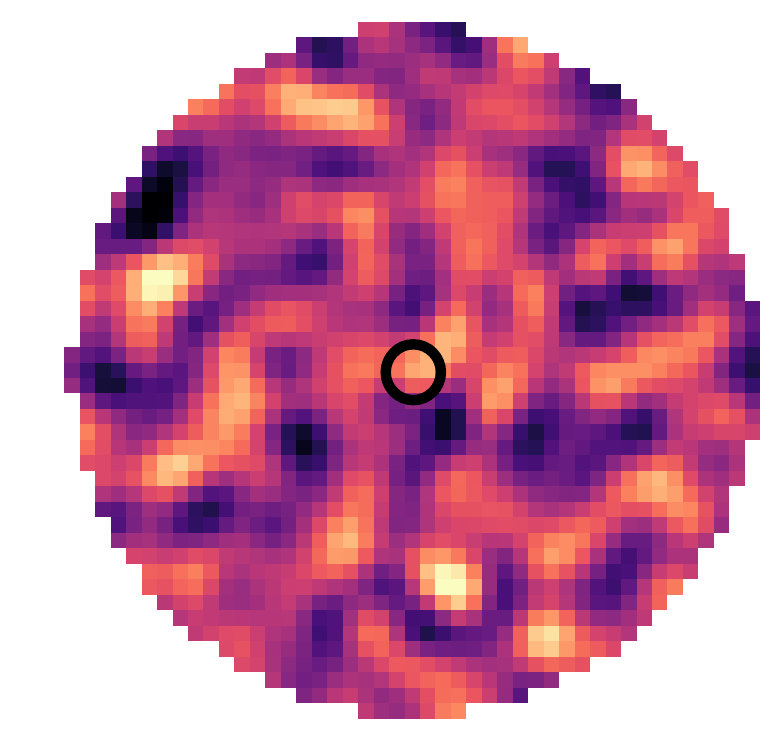}}
	\subfigure[ULASJ1455+1230]{\includegraphics[trim= 120 70 115 65 ,clip,width=0.23\textwidth]{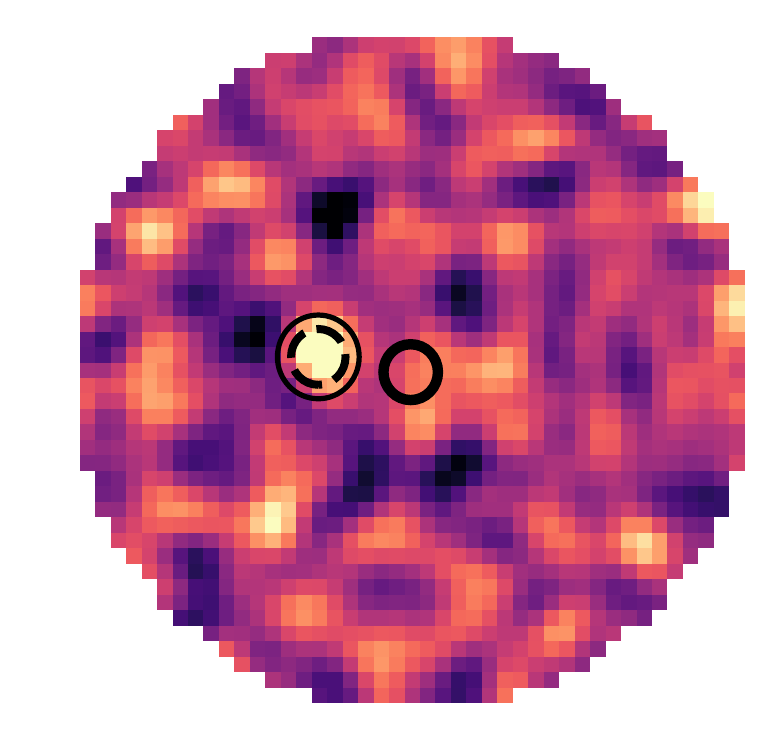}}
    \subfigure[ULASJ1539+0057]{\includegraphics[trim= 120 70 115 65 ,clip,width=0.23\textwidth]{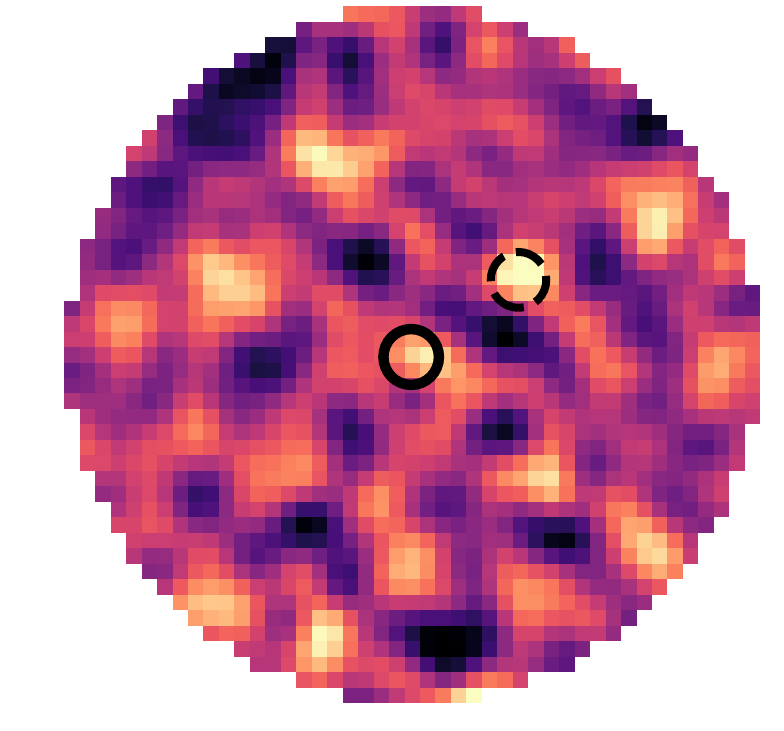}}
	\subfigure[VHSJ1556-0835]{\includegraphics[trim= 120 70 115 65 ,clip,width=0.23\textwidth]{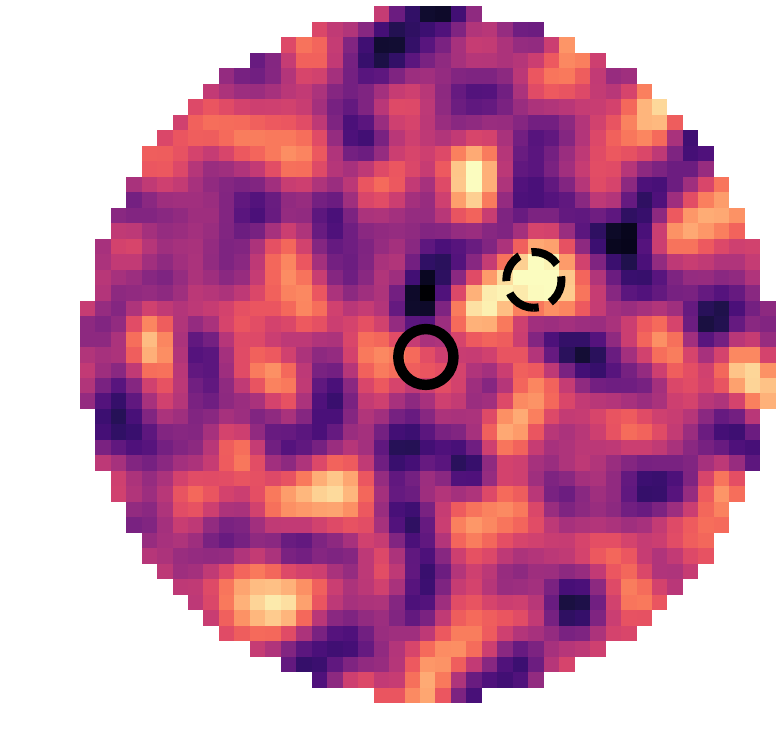}}
    \subfigure[VHSJ2109-0026]{\includegraphics[trim= 120 70 115 65 ,clip,width=0.23\textwidth]{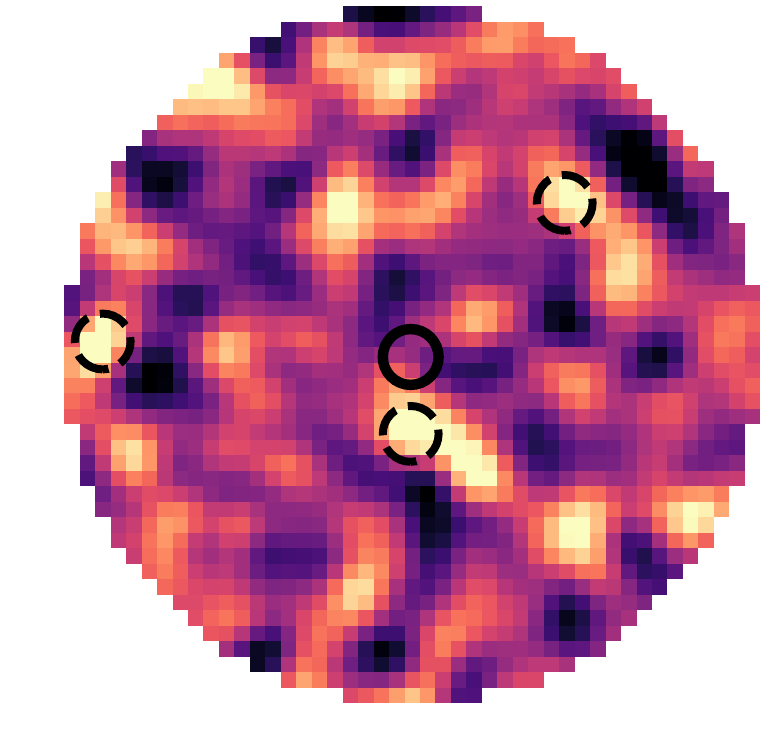}}
    \subfigure[VHSJ2143-0643]{\includegraphics[trim= 120 70 115 65 ,clip,width=0.23\textwidth]{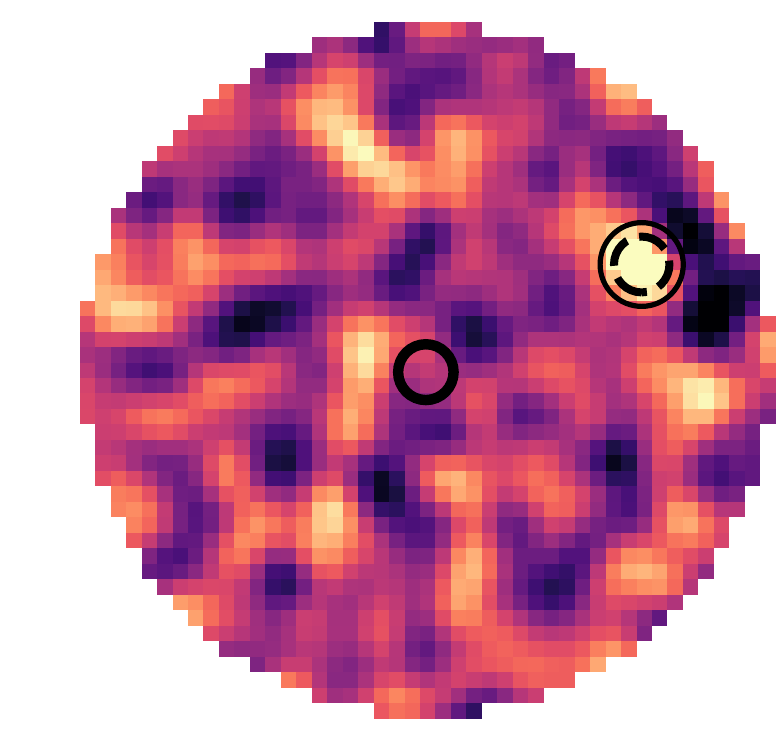}}
	\subfigure[VHSJ2144-0523]{\includegraphics[trim= 120 70 115 65 ,clip,width=0.23\textwidth]{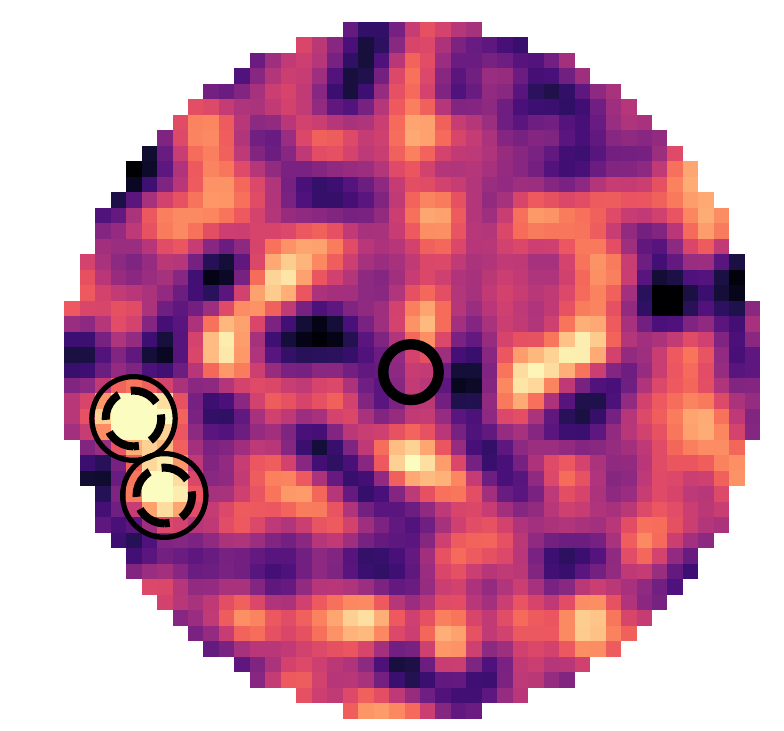}}
	\subfigure[ULASJ2200+0056]{\includegraphics[trim= 120 70 115 65 ,clip,width=0.23\textwidth]{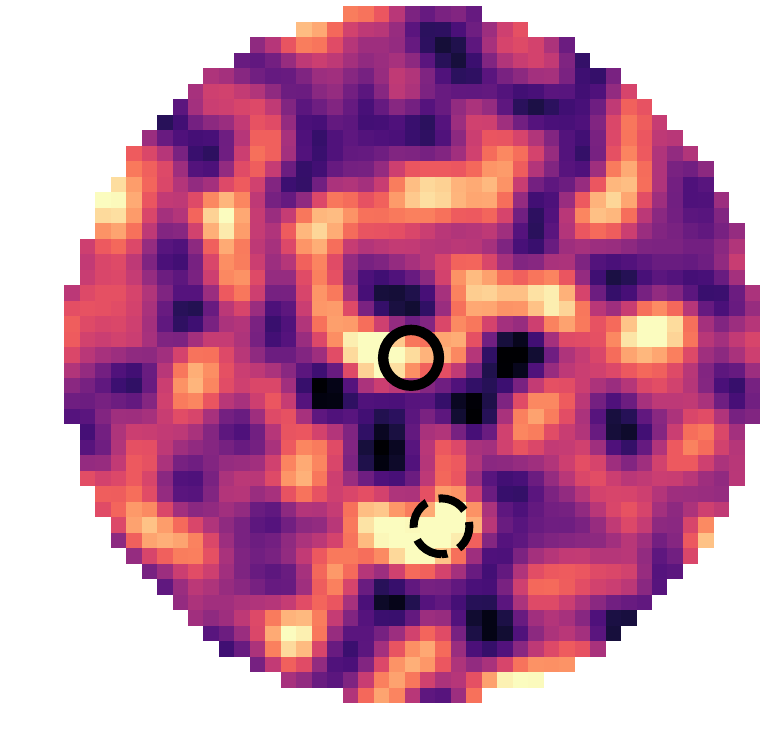}}
    \subfigure[ULASJ2224-0015]{\includegraphics[trim= 120 70 115 65 ,clip,width=0.23\textwidth]{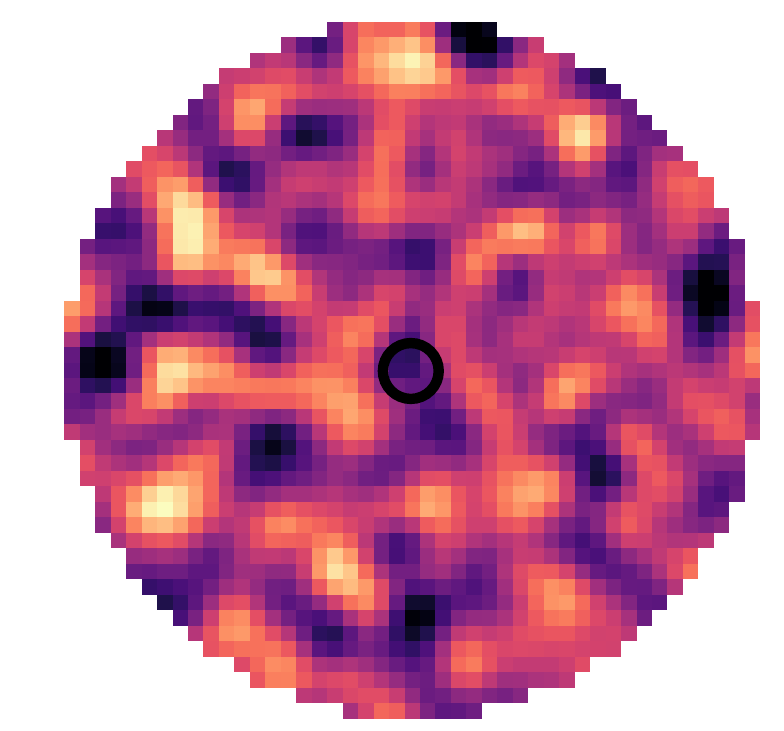}}
	\subfigure[ULASJ2315+0143]{\includegraphics[trim= 120 70 115 65 ,clip,width=0.23\textwidth]{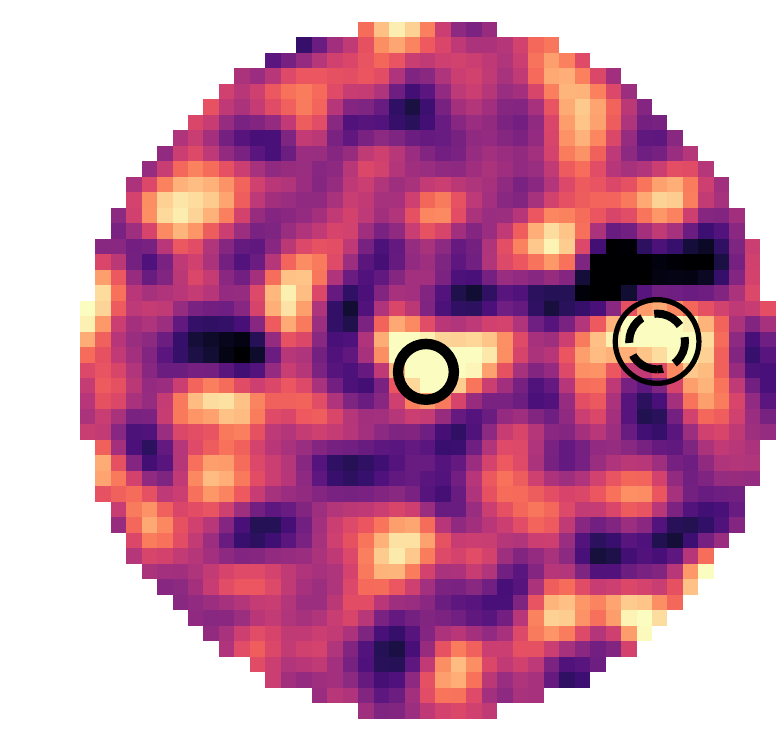}}
	\subfigure[VHSJ2355-0011]{\includegraphics[trim= 120 70 115 65 ,clip,width=0.23\textwidth]{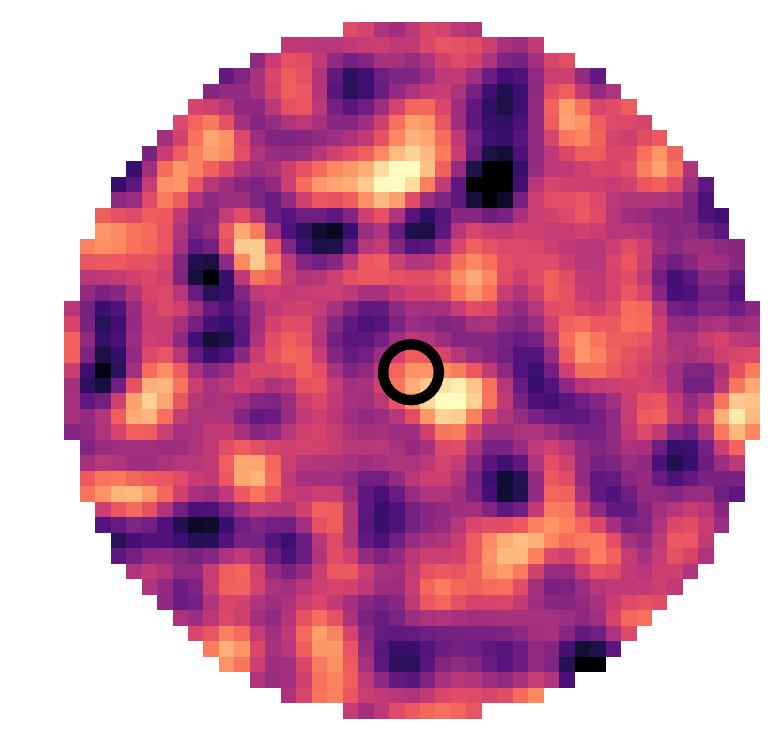}}
\caption{Three-arcminute diameter maps for all quasar targets, showing the locations of serendipitous sources detected at $>$3.5$\sigma$ (\emph{dotted}) and $>$4$\sigma$ (\emph{solid}). Central aperture denotes the 14.6 arcsec diameter SCUBA-2 beam at the position of each quasar in our sample. North is up, East is to the left.}
\label{fig:detect_all}
\end{figure*}

Table~\ref{tab:serendipitous} shows the number of serendipitous detections found above 3, 3.5 and 4$\sigma$ across our full sample, along with the corresponding number of negative detections for the same detection threshold. Although some negative sources are expected due to `bowling' around real positive sources in the SCUBA-2 images, we find the number of these negative sources to make up half of the positive sources detected at $>$3$\sigma$, meaning that only half of these positive sources are likely to be real. At 3.5$\sigma$, the number of negative sources drops to 38 per cent of the positive detections, with no maps containing multiple negative detections. When the detection threshold is further increased to 4$\sigma$, the negative sources account for just 25 per cent of the positive detections, meaning the majority of sources detected above this threshold are unlikely to be artefacts of the noise in the image. Fig~\ref{fig:detect_all} shows the location of all 3.5 and 4$\sigma$ 850$\mu$m detections in all 19 of our SCUBA-2 maps.

\begin{table*}
	\centering
	\caption{Number counts for positive and negative sources lying within a 1.5 arcmin radius of each quasar target, detected with a significance above 3, 3.5 and 4$\sigma$.}
	\label{tab:serendipitous}
	\begin{tabular}{lcccccc}
		\hline
		Name & >3.0$\sigma$ & <-3.0$\sigma$ & >3.5$\sigma$ & <-3.5$\sigma$ & >4.0$\sigma$ & <-4.0$\sigma$\\
		\hline
		ULASJ0016-0038 &  3 & 2 & 2	& 1	& 0	& 0 \\	
		ULASJ0041-0021 &  4 & 2 & 0	& 1	& 0	& 1 \\
		VHSJ1117-1528  &  1 & 0 & 1	& 0	& 1	& 0 \\
		VHSJ1122-1919  &  5 & 2 & 3	& 1	& 2	& 0 \\
		ULASJ1216-0313 &  3 & 2 & 2	& 1	& 2	& 0 \\
		ULASJ1234+0907 &  1 & 0 & 1	& 0	& 0	& 0 \\
		VHSJ1301-1624  &  3 & 0 & 1	& 0	& 0	& 0 \\
		VHSJ1350-0503  &  1 & 2 & 1	& 0	& 1	& 0 \\
		VHSJ1409-0830  &  2 & 1 & 0	& 0	& 0	& 0 \\
		ULASJ1455+1230 &  1 & 1 & 1	& 1	& 1	& 0 \\
		ULASJ1539+0057 &  1 & 1 & 1	& 0	& 0	& 0 \\
		VHSJ1556-0835  &  2 & 0 & 1	& 0	& 0	& 0 \\
		VHSJ2109-0026  &  5 & 2 & 3	& 1	& 0	& 0 \\
		VHSJ2143-0643  &  1 & 2 & 1	& 1	& 1	& 0 \\
		VHSJ2144-0523  &  2 & 1 & 2	& 0	& 2	& 0 \\
		ULASJ2200+0056 &  3 & 1 & 2	& 0	& 0	& 0 \\
		ULASJ2224-0015 &  0 & 2 & 0	& 0	& 0	& 0 \\
		ULASJ2315+0143 &  3 & 2 & 2	& 1	& 2	& 1 \\
		VHSJ2355-0011  &  1 & 1 & 0	& 1	& 0	& 1 \\
		\hline  
		TOTAL & 42 & 24 & 24 & 9 & 12 & 3 \\                                                          
	\end{tabular}
\end{table*}

\begin{figure*}
	\centering  
    \subfigure[$>$3.5$\sigma$]{\includegraphics[trim= 40 20 0 40 ,clip,width=0.45\textwidth]{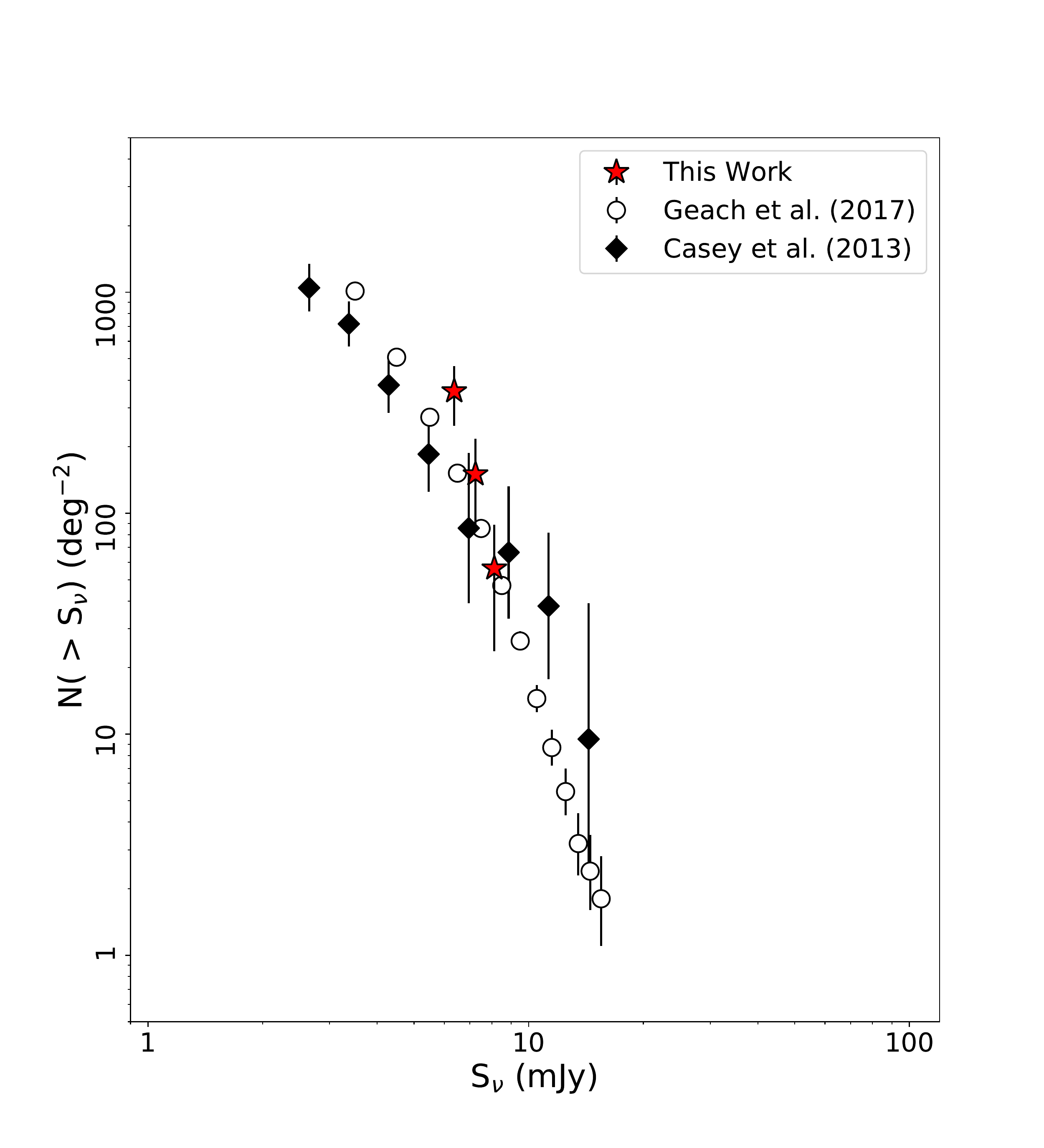}}
    \subfigure[$>$4.0$\sigma$]{\includegraphics[trim= 40 20 0 40 ,clip,width=0.45\textwidth]{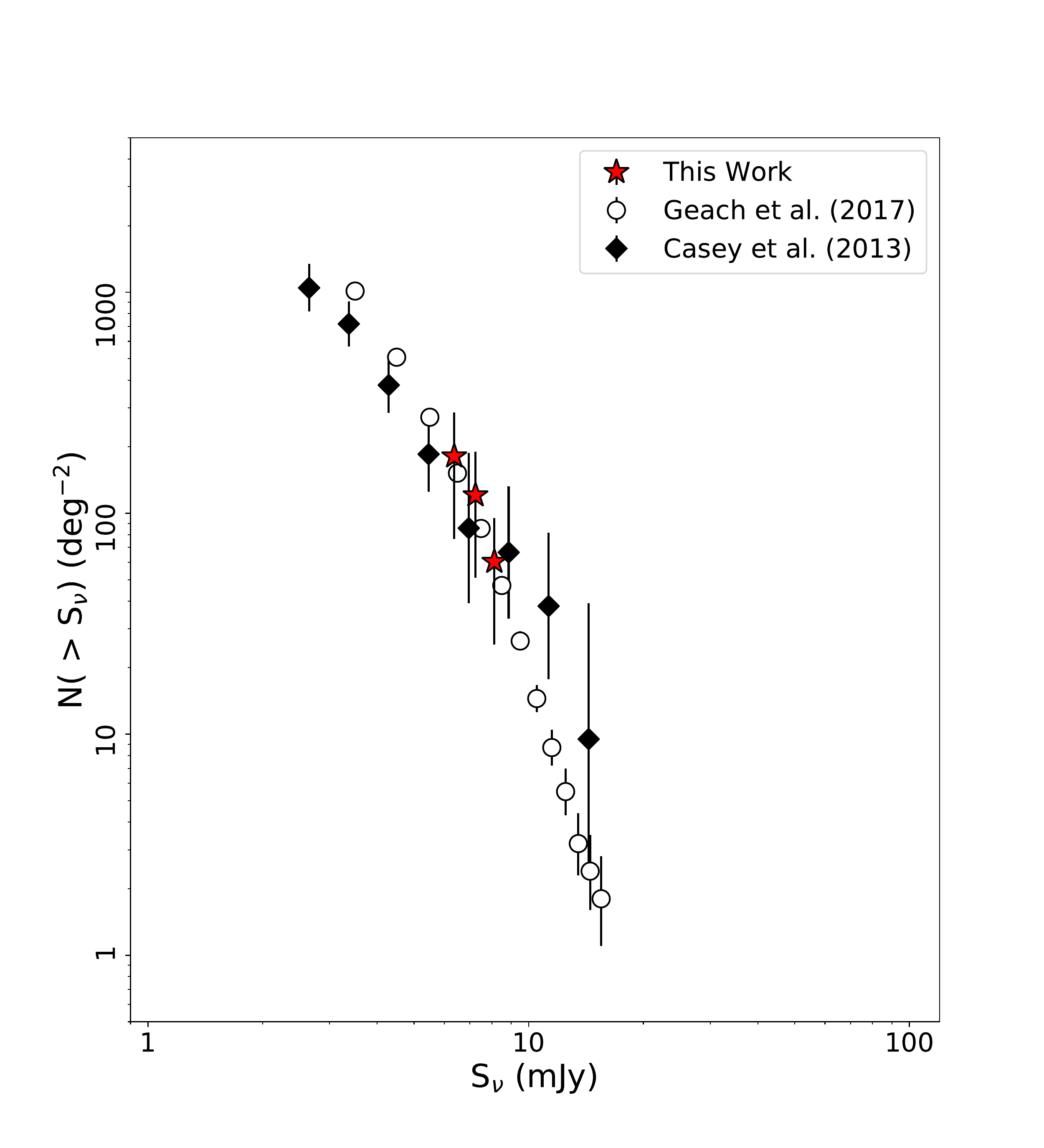}}
\caption{$>$3.5$\sigma$ (a) and $>$4$\sigma$ (b) detections within 1.5 arcmin radius of the reddened quasar targets (red stars) are compared to the blank-field counts from \citet{geach17} (open circles) and \citet{casey13} (solid diamonds). Number counts for our sample have been corrected to represent only the fraction of `real' sources expected based on the statistics provided in Table~\ref{tab:serendipitous}.} 
\label{fig:sources}
\end{figure*}

In total, we find 24 (12) sub-mm sources at $>$3.5$\sigma$ ($>$4$\sigma$) across our reddened quasar sample, corresponding to a total survey area of 134.3 arcmin$^2$. To test whether this is indicative of an overdensity, the number counts for the $>$3.5$\sigma$ ($\gtrsim$5.6 mJy) and $>$4$\sigma$ ($\gtrsim$6.4 mJy) detections are directly compared to sub-mm blank field samples \citep{geach17} by plotting the cumulative frequency of the source counts as a function of their 850$\mu$m flux (Fig.~\ref{fig:sources}). Using the number counts of the positive and negative sources in Table~\ref{tab:serendipitous}, we expect 62 per cent of the 3.5$\sigma$ and 75 per cent of the 4$\sigma$ detections to be real. We account for this when comparing to the sub-mm blank fields in Fig.~\ref{fig:sources} by scaling the number of sources by the fraction we predict to be real. 

Fig.~\ref{fig:sources} shows that at both $>$3.5 and $>$4$\sigma$ the source counts in our reddened quasar survey are entirely consistent with those in the blank fields when averaging over the full survey area for our sample of 19 reddened quasars. However, there are clearly individual fields where we do see some evidence for over-densities. For example, above our 3.5$\sigma$ flux-limit of $\simeq$5.6 mJy we would predict 0.5 sources within a single SCUBA-2 map of radius 1.5 arcmin based on the \citet{geach17} blank-field counts. There are four quasars where we see two or three sources in the SCUBA-2 maps brighter than this flux-limit (Table \ref{tab:serendipitous}). Even accounting for the false detection rate of 38 per cent at this threshold, there are still 1-2 real sources detected around the quasar in these fields, a factor of 2-4 higher than predicted from the blank-field counts. It is thus reasonable to conclude that while the majority of our reddened quasars do not appear to reside in over-dense regions, there are nevertheless examples of heavily reddened quasars that may do. It should also be noted that any sub-mm sources $\lesssim$30 arcsec away from the quasar will not be resolved as a separate source in the SCUBA-2 images and may be blended with the quasar emission. As discussed in the following sub-section, higher spatial resolution Atacama Large Millimetre Array (ALMA) observations for two quasars confirm the presence of close companions. 

\subsubsection{Multiplicity from ALMA}
\label{sec:alma}

\begin{figure}
	\centering  
    \subfigure[ULASJ1234+0907]{\includegraphics[trim= 130 0 120 0 ,clip,width=0.23\textwidth]{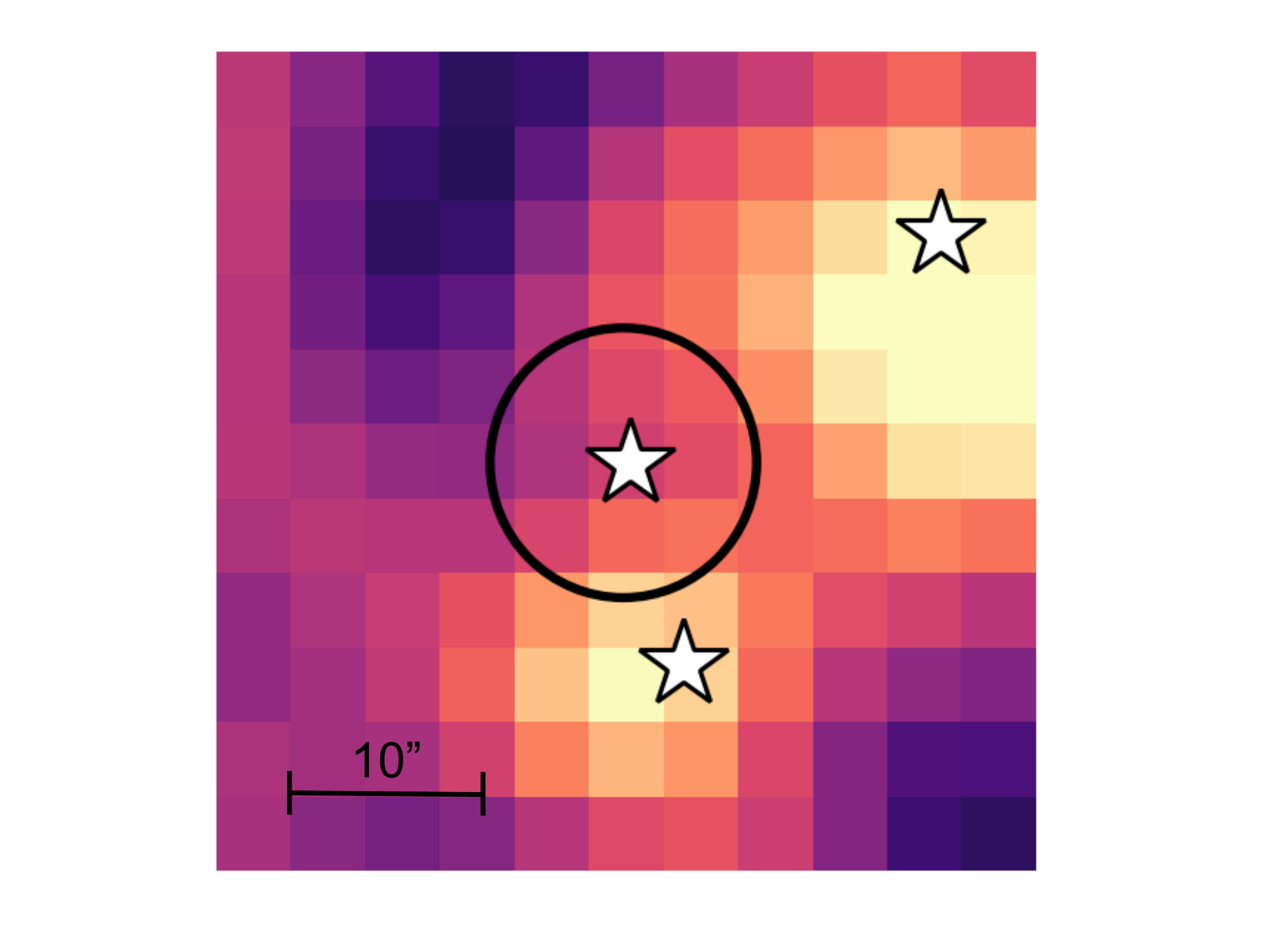}}
    \subfigure[ULASJ2315+0143]{\includegraphics[trim= 120 0 120 0 ,clip,width=0.23\textwidth]{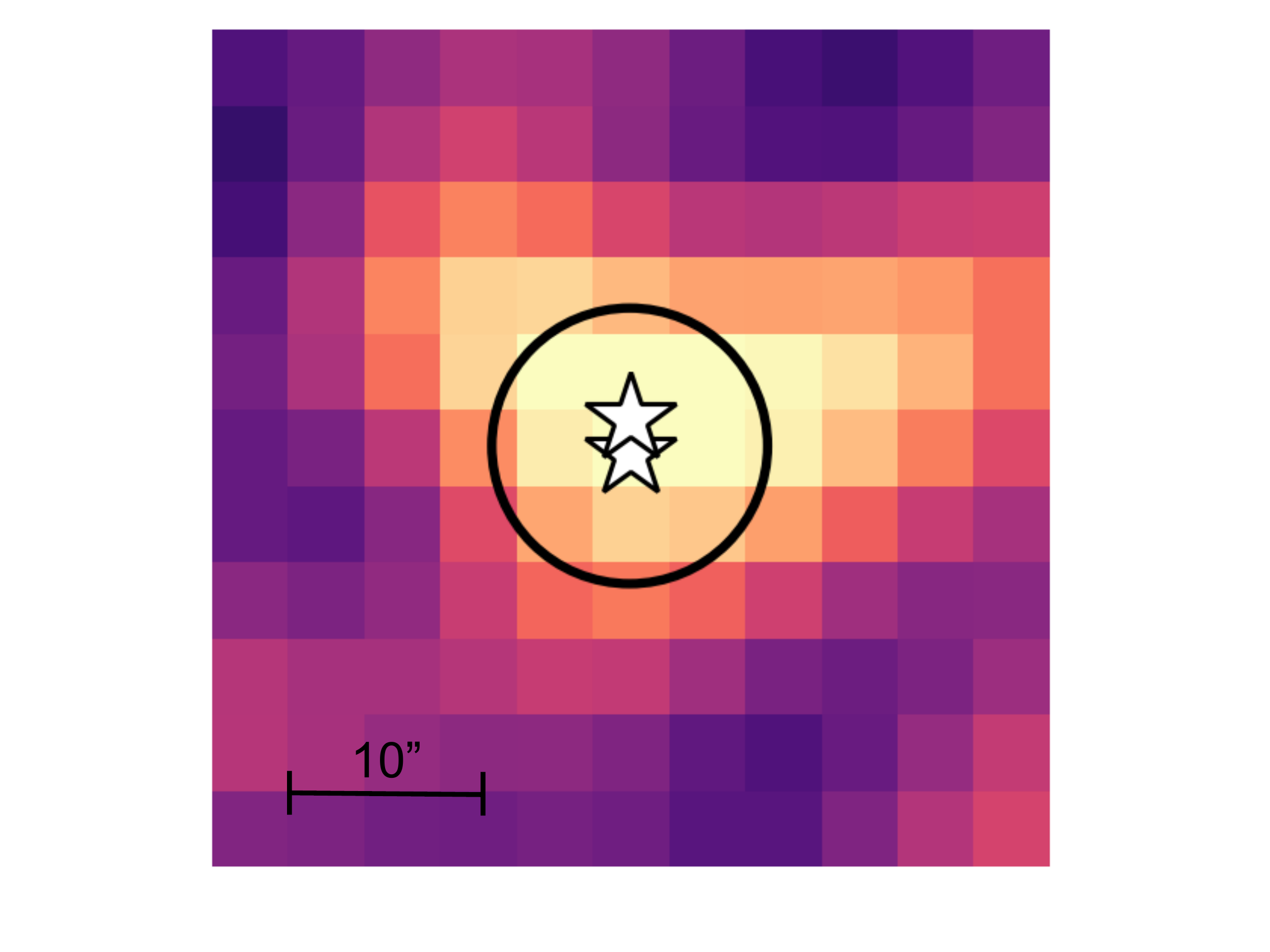}}
\caption{SCUBA2 maps for the two sources in our sample also observed with ALMA -- ULASJ1234+0907 and ULASJ2315+0143. White stars denote the location of individual sources detected with higher-resolution ALMA data \citep{banerji17}. North is up, East is to the left.}
\label{fig:alma}
\end{figure}

Two of the reddened quasars in our sample (ULASJ1234+0907 and ULASJ2315+0143) have also been observed with ALMA \citep{banerji17,banerji18} in Band 3 and Band 6, which trace the 1.24mm and 2.92mm continuum emission. With the higher resolution of the ALMA observations ($\simeq$2-3 arcsec compared to $\simeq$15 arcsec with SCUBA-2), \citet{banerji18} find that both ULASJ1234+0907 and ULASJ2315+0143 have multiple sources present near the quasar. The locations of the ALMA continuum detected sources are overlaid on the SCUBA-2 850$\mu$m maps in Fig.~\ref{fig:alma}. With ALMA, \citet{banerji18} also detect multiple molecular emission lines from these sources and in both ULASJ1234+0907 and ULASJ2315+0143. The additional ALMA sources marked in Fig.~\ref{fig:alma} are found to lie at a similar redshift to the quasar, confirming that they are part of the same overdensity. This provides direct evidence that at least some fraction of the reddened quasars in our sample reside in high density sub-mm environments. We note that ULASJ1234+0907 is undetected with SCUBA-2, lying below the 3$\sigma$ detection limit of our sample. However, the two companion galaxies near the quasar appear to have higher 850$\mu$m flux densities (7.48$\pm$2.03 mJy and 5.92$\pm$2.03 mJy) than the quasar. 

Fig.~\ref{fig:alma} shows that both the ALMA sources detected in the map of ULASJ2315+0143 lie within the beamsize of SCUBA-2, so the SCUBA-2 850$\mu$m flux density represents the sum of the two ALMA sources. The same may also be true for the other two heavily reddened quasars detected at 850$\mu$m. Indeed, in the case of ULASJ2200+0056, the 850$\mu$m peak is offset from the quasar position, which could indicate that the bulk of the sub-mm emission is not associated with the quasar but rather a companion galaxy. 

\section{Discussion}
\label{sec:discussion}

In order to put our new 850$\mu$m observations of the heavily reddened quasar population at $z\sim2$ in context, we compare the results to similar observations of other high-redshift, high-luminosity quasars. In particular, we consider a sample of UV-luminous, essentially unobscured, quasars observed with SCUBA \citep{priddey03} as well as the population of mid-infrared luminous Hot Dust Obscured Galaxies (HotDOGS; \citealt{eisenhardt12}) discovered using the \textit{WISE} All-Sky Survey, which have considerably higher line of sight extinctions compared to our reddened quasars. 

\subsection{UV-Luminous Unobscured Quasars}

\begin{figure}
	\centering
	\includegraphics[trim= 0 0 35 20 ,clip,width=0.45\textwidth]{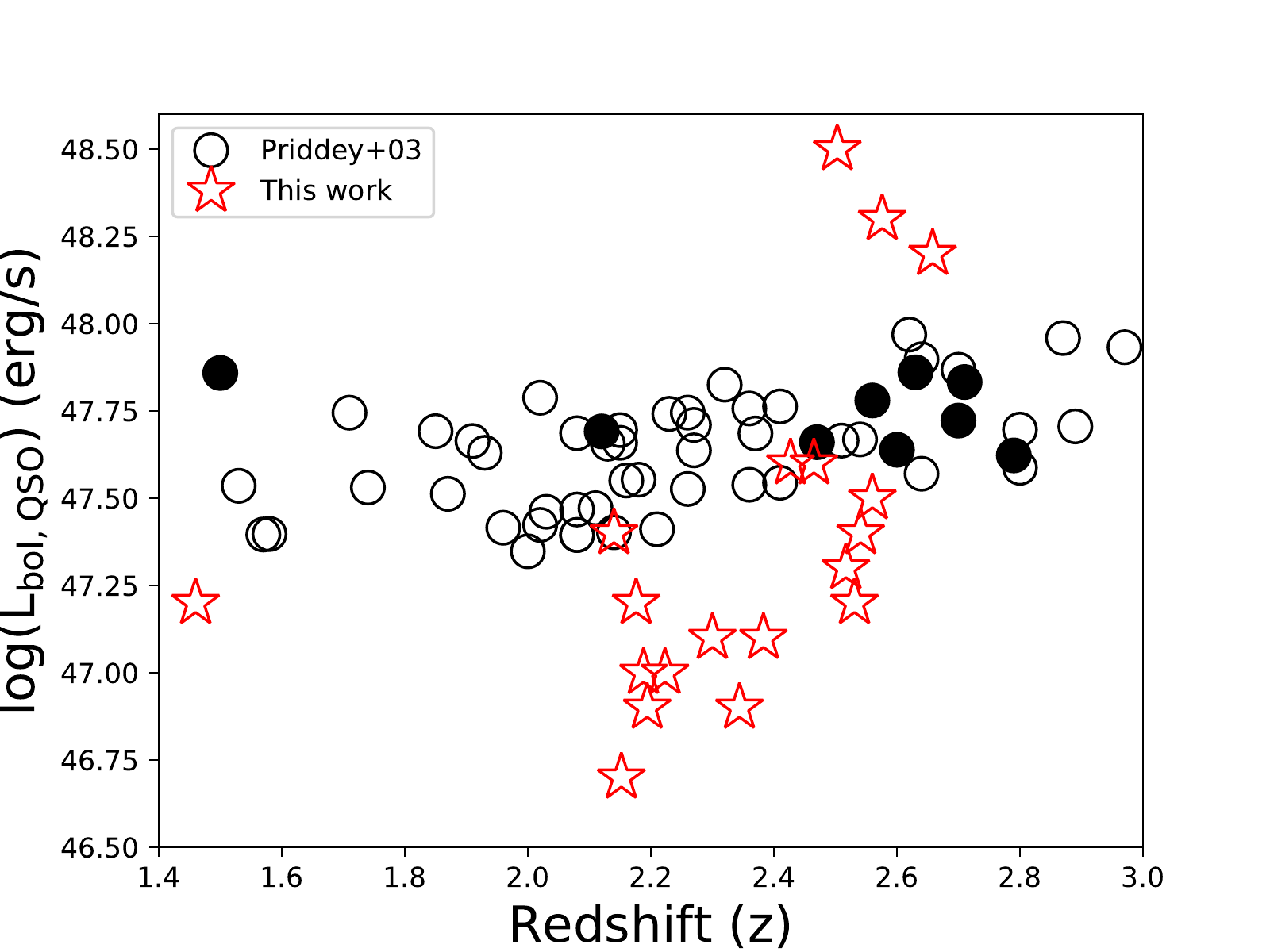}
\caption{A comparison of the bolometric quasar luminosities (L$_{\rm{bol}}$) for our sample (\textit{red stars}) and the UV-luminous quasar sample of \citet{priddey03} (\textit{black circles}). Filled symbols denote all sources detected above a flux limit $>$10mJy, corresponding to the 3$\sigma$ detection threshold in \citet{priddey03}. The quasar luminosities in our sample have been corrected for the effects of dust.}
\label{fig:pridd_comp}
\end{figure}

\citet{priddey03} conducted a large 850$\mu$m survey of 57 UV-luminous quasars at $z\sim2$ with SCUBA down to a relatively bright 3$\sigma$ flux-limit of 10 mJy. Only nine of the 57 quasars were detected at 850$\mu$m in their study. In Fig.~\ref{fig:pridd_comp} we compare the quasar bolometric luminosity and redshift distribution of our sample with the UV-bright quasars from \citet{priddey03}. In the case of our sample, L$_{\rm{bol}}$ is derived from the dust-corrected flux at 5100\AA\, after applying a bolometric correction factor of 7 \citep{vestergaard06}. While the redshift range of our quasars overlaps with the UV-bright quasars, overall our dusty quasars are intrinsically fainter than the UV-luminous quasars. Our 850$\mu$m survey also probes a factor of two deeper than the SCUBA survey, reaching a 3$\sigma$ flux-limit of $\sim$4.8 mJy. 

Nine of the 57 UV-luminous quasars presented in \citet{priddey03} are brighter than 10mJy, corresponding to 16 per cent of their full sample. By comparison, none of our heavily reddened quasars are detected above this flux threshold, although ULASJ1216-0313 has a flux-density of 9.9$\pm$1.8 mJy, consistent with a $>$10mJy flux given the 1$\sigma$ uncertainties. Overall therefore it appears that a smaller fraction of our quasar sample is extremely sub-mm bright compared to the UV-luminous quasars in \citet{priddey03}. We note however that the median dust-corrected L$_{\rm{bol}}$ of our sample is 10$^{47.2}$ ergs$^{-1}$, compared to 10$^{47.7}$ ergs$^{-1}$ in the sample of \cite{priddey03}. The lower sub-mm detection rate for quasars in our sample could therefore potentially be explained by their lower luminosities. If we only consider quasars with L$_{\rm{bol}} > 10^{47.25}$ ergs$^{-1}$, which includes the entire \citet{priddey03} sample, the 850$\mu$m detection rates for the two samples are consistent given the small number of detections. 

\subsection{Mid-Infrared Luminous HotDOGs}

\begin{figure}
	\centering
	\includegraphics[trim= 0 0 35 20 ,clip,width=0.45\textwidth]{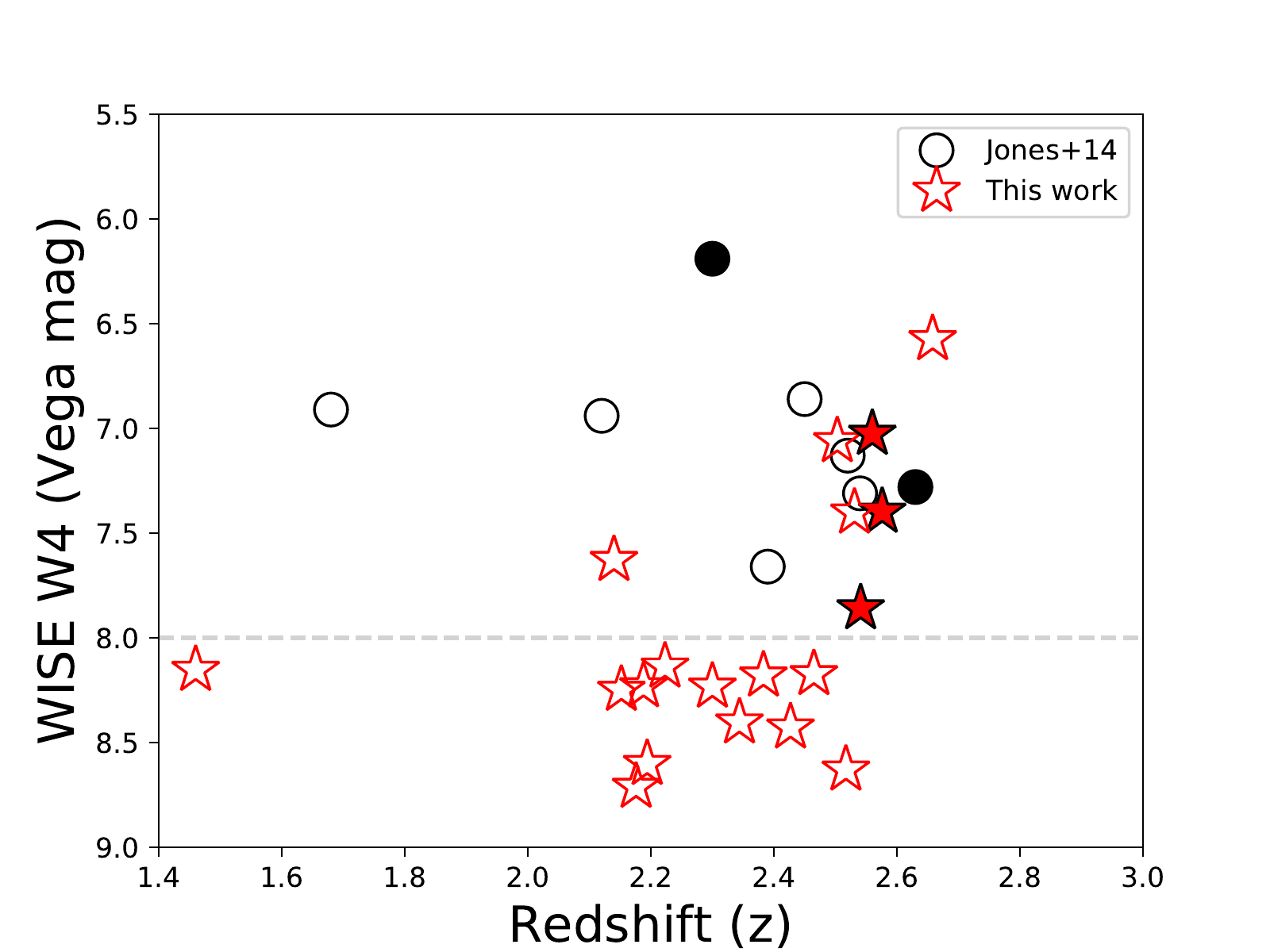}
\caption{A comparison of the WISE W4 magnitudes (Vega) of our quasar sample (\textit{red stars}) and the HotDOGs from \citet{jones14} (\textit{black circles}). Filled symbols denote all sources detected above a flux limit $>$5.8 mJy, corresponding to the 3$\sigma$ detection threshold in our sample. The HotDOG W4 magnitude-selection cut at 8\,mag is also shown (\textit{grey line})}
\label{fig:jones_comp}
\end{figure}

\cite{jones14} outline a sample of ten HotDOGs, selected from the \textit{WISE} All-Sky Survey, with dust extinctions of A$_{V}$ $\gtrsim$ 15 mag towards the quasar continuum -- significantly larger than the A$_{V}$ $\simeq$ 2-6 mag values for our sample. Of the ten HotDOGs \citep{jones14}, eight have spectroscopic redshifts within the range of our reddened quasar sample (1.4 $<$ z $<$ 3) and therefore form our comparison sample. In order to make a direct comparison to the reddened quasar population, we select only the seven quasars in our sample with 22$\mu$m (W4) magnitudes brighter than 8.0 (Vega), above which the entire \cite{jones14} sample lies. The redshifts and 22$\mu$m (W4) magnitudes for the two samples are presented in Fig.~\ref{fig:jones_comp}.

Two of the eight HotDOGs (25 per cent) are detected at an 850$\mu$m flux-limit brighter than 5.8 mJy, corresponding to the 3$\sigma$ detection threshold used in this paper. By comparison, three of the seven heavily reddened quasars (43 per cent) with W4 magnitudes brighter than 8.0 (Vega) are detected at 850$\mu$m above this flux threshold. Furthermore, a second sample of 30 HotDOGs is outlined by \cite{jones15}, in which targets are additionally required to be detected as compact radio sources. Although only four of these radio-selected HotDOGs have spectroscopic redshifts within the range of our reddened quasar sample, we note that none are detected $>$5.8 mJy at 850$\mu$m. Given the small numbers of detections however, there is no statistically significant evidence for a difference in the 850$\mu$m detection rates between HotDOGs and reddened quasars. 

In Section~\ref{sec:environments} we concluded that our heavily reddened quasars do not, on average, reside in over-dense regions of the universe although some individual quasars may do. Similar studies have been conducted for the HotDOG populations in \citet{jones14, jones15}, who find evidence for over-densities around their AGN. The over-densities appear to be more pronounced in the radio-selected HotDOG population. However, many of the radio-HotDOGs from \citet{jones15} lack spectroscopic redshifts and thus it is difficult to make a more direct comparison using luminosity- and redshift-matched samples.

\section{Conclusions}
\label{sec:conclusions}

We have presented new 850$\mu$m SCUBA-2 observations for a sample of 19 luminous, heavily reddened quasars (A$_{\rm{V}}$ $\simeq$ 2-6 mags) at 1.4 $\leq$ z $\leq$ 2.7 -- the peak epoch of both star formation and black hole accretion.

(i) Three of the 19 quasars are detected at $>$3$\sigma$ significance corresponding to 850$\mu$m flux densities of $>$4.8 mJy. Assuming the 850$\mu$m emission is dominated by cool dust heated by star formation in the quasar host galaxies, implies very high SFRs of $\sim$2500 - 4500 M$_{\rm{\odot}}$yr$^{-1}$ in the quasar host galaxies. Even when assuming a significant contribution to the 850$\mu$m flux from the quasar itself, very high star formation rates of $\sim$600 - 1500 M$_{\rm{\odot}}$yr$^{-1}$ are nevertheless inferred for two of the three detected quasars. We stack the remaining 16 undetected quasars to obtain a 3$\sigma$ upper limit on the SFR of $<$880 M$_\odot$ yr$^{-1}$ for these quasars. 

(ii) Several serendipitous sub-mm detections are found in the 1.5 arcmin radius 850$\mu$m maps around each quasar. Considering the 19 fields as a whole, the number counts of these serendipitous detections in the full survey area are consistent with those from blank-field observations. However, several of the individual quasars show evidence for residing in over-dense regions, a conclusion supported by higher spatial-resolution ALMA observations of several of the quasars. 

(iii) We compare the 850$\mu$m detection rate of our heavily reddened quasars to both UV-luminous unobscured quasars and the much more highly obscured population of mid-infrared luminous radio-detected HotDOGs. Given the relatively bright flux limits of the 850$\mu$m surveys that have been conducted for these populations, the number of detected quasars in each sample is small and there is overall no significant evidence for a difference in 850$\mu$m properties. 

(iv) The mid-infrared luminous HotDOGs, particularly those selected to host compact radio sources, do however appear to reside in more over-dense regions of the universe compared to our heavily reddened quasars. However, as many of the HotDOGs lack spectroscopic redshifts, further observations are required to confirm the apparent difference. 

Given the relatively low detection rate of these heavily reddened quasars with SCUBA-2, a more sensitive dust-continuum survey, e.g. with ALMA, is required to better understand the cool-dust emission properties of the population. ALMA observations would further allow us to obtain spectroscopic identifications for the sub-mm bright sources found around the quasars in the SCUBA-2 maps, and therefore to determine unambiguously whether they are part of the same over-density as the quasar. 

\section*{Acknowledgements}

We thank the anonymous referee for a constructive review of the paper. CFW acknowledges funding via an STFC studentship. MB acknowledges funding from The Royal Society via a University Research Fellowship. PCH acknowledges funding from STFC via the Institute of Astronomy, Cambridge, Consolidated Grant. We thank Myrto Symeonidis for providing the quasar template from \citet{symeonidis16}. Based on observations conducted on the James Clerk Maxwell Telescope as part of Program M17AP008 (PI:Banerji). The James Clerk Maxwell Telescope is operated by the East Asian Observatory on behalf of The National Astronomical Observatory of Japan; Academia Sinica Institute of Astronomy and Astrophysics; the Korea Astronomy and Space Science Institute; Center for Astronomical Mega-Science (as well as the National Key R\&D Program of China with No. 2017YFA0402700). Additional funding support is provided by the Science and Technology Facilities Council of the United Kingdom and participating universities in the United Kingdom and Canada. Additional funds for the construction of SCUBA-2 were provided by the Canada Foundation for Innovation.




\bibliographystyle{mnras}
\bibliography{references} 








\bsp	
\label{lastpage}
\end{document}